\documentclass[aps]{revtex4-1}

\usepackage{amssymb}
\usepackage{amsmath}

\begin{document}

\title{Current Status of Nuclear Physics Research}

\author{ Carlos A. Bertulani$^a$ and Mahir S. Hussein$^{b,c}$}

\address{$^a$ Department of Physics and Astronomy, Texas A\&M University-Commerce,
                                   Commerce TX 75429, USA}
\address{$^b$ Instituto de Estudos Avan\c{c}ados and Instituto de F\'{\i}sica, Universidade de S\~ao Paulo, 
                                     05314-970 S\~ao Paulo, SP, Brazil}                                                                  
\address{$^c$Departamento de F\'{\i}sica, Instituto Tecnol\'ogico de Aeron\'autica, DCTA,
                               12.228-900 S\~ao Jos\'e dos Campos, SP, Brazil}

\date{\today}

\begin{abstract}
In this review we discuss the current status of research in nuclear physics which is being carried out in different centers in the World. For this purpose we supply a short account of the development in the area which evolved over the last 9 decades, since the discovery of the neutron. The evolution of the physics of the atomic nucleus went through many stages  as more data become available. We briefly discuss models  introduced to discern the physics behind the experimental discoveries, such as the shell model, the collective model, the statistical model, the interacting boson model, etc., some of these models may be seemingly in conflict with each other, but this was shown to be only apparent. 

The richness of the ideas and abundance of theoretical models attests to the important fact that the nucleus is a really singular system in the sense that it evolves from two-body bound states such as the deuteron, to few-body bound states, such as $^4$He, $^7$Li, $^9$Be etc. and up the ladder to heavier bound nuclei containing up to more than 200 nucleons. Clearly statistical mechanics, usually employed in systems with very large number of particles, would seemingly not work for such finite systems as the nuclei, neither do other theories which are applicable to condensed matter. The richness of nuclear physics stems from these restrictions. New theories and models are presently being developed. Theories of the structure and reactions of neutron-rich and proton-rich nuclei, called exotic nuclei, halo nuclei, or Borromean nuclei, deal with the wealth of experimental data that became available in the last 35 years. Further, nuclear astrophysics and stellar and Big Bang nucleosynthesis have become a more mature subject. Due to limited space, this review only covers a few selected  topics, mainly those with which the authors have worked on.
 Our aimed potential readers of this review are nuclear physicists and physicists in other areas, as well as graduate students interested in pursuing a career in nuclear physics. 
\end{abstract}

\pacs{20., 21., 23., 24., 25., 26.}

\maketitle

\section{Nucleons, Nuclear Structure, and Politics}

Nuclear Physics evolved in a rather logical manner; from simple ordered {\it shell model} tailored in accordance with atomic structure to reproduce the magic numbers \cite{Ma49,Ha49}, to collective models needed to describe vibrations and cases involving non-spherical nuclei. The collective model is now considered an elaborate extension of the shell model through the inclusion of nucleon-nucleon correlations, which are predominantly of the pairing plus quadrupole type \cite{BM69}. 
In fact the shorter-range pairing correlations were introduced into nuclear physics at almost the same time that the BCS theory of {\it superconductivity} was developed \cite{BCS}. The quadrupole-quadrupole interaction is responsible for long-range correlations and gives rise to deformed mean field necessary in the description of deformed nuclei with their vibrational and rotational spectra \cite{BM69}. For general texts on nuclear physics theory we suggest Refs.  \cite{BM69,Weisskopf,Sat83,Feshbach1993,Tal93,wong,BD04,RS04,Bertu07,CH13}

Correlations involving particles and  holes are also used in the description of excited states in nuclei, the most notable of these is the so-called {\it giant resonances}, similar to plasmon resonances in atom clusters \cite{BK47,SJ50,GT48,Berm75,Bert76,Spet81}. In the 1970's algebraic models based on group theory were introduced, guided by the pairing plus quadrupole models alluded to above. The resulting {\it Interacting Boson Model} (IBM) has as building blocks pairs of correlated particles (neutrons or protons) coupled to angular momentum 0 and 2 and treated as bosons \cite{AI75,AI81,Tal93}. The resulting Hamiltonian can be analyzed using symmetry concepts and the corresponding spectrum is obtained analytically with the help of group theory. 

With the development of models of nuclear structure, which pointed out certain degrees of freedom to be studied to test these models,  there was the accompanying development of a theory of {\it nuclear reactions}, aimed to probe these different degrees of freedoms. The concepts of {\it compound nucleus} \cite{Bohr,Fes58a,KRY1963,BF1963,FKL1967,Feshbach1993}  and {\it direct reactions} \cite{Sat83,BD04} became a common language with the bombarding energy of the probing projectile being the deciding factor in applying either one of the two concepts: low energy projectiles are captured by the target to form a highly excited compound nucleus which then decays into the different open channels, while higher energy projectiles excites the target nucleus or transfer a nucleon or more \cite{Feshbach1993}. A unifying concept that deals with both compound and direct processes in an average way is the optical model \cite{Fes58a}, where the scattering Schr\"odinger equation is solved with  a complex potential whose imaginary part is meant to simulate processes that remove flux from the elastic channel described by this same equation. It is this rich variety of possibilities which prompted the great advances made in the quantum scattering theory that underlines nuclear reactions. In this review we give a brief account of these development. 

As decades passed by, the main interests in nuclear physics were broadened and strong overlaps with other areas of science emerged. Our knowledge about the nucleons also changed with time. Now we know that they are built by {\it deceiving confined particles}; the quarks and gluons, generically known as {\it partons}. New theories and experiments are now dedicated to probe the sub-nucleon degrees of freedom and how they change in the nuclear medium, in stars and in the evolution of our Universe. We are witnessing an increasing enthusiasm in the nuclear physics community with its stronger involvement in problems with interface with many other areas, such as Chemistry, Biology, Cosmology, Astrophysics, Energy, Medicine, etc. 

Needless to mention that Nuclear Physics carries a big burden of being tightly involved with political issues such as national security and non-proliferation of nuclear weapons, a serious threat to the survival of the human race.  But, it is also rewarding to witness the involvement of many nuclear physicists in politics, a good example being our colleague  E. Moniz, a nuclear theorist and since 2013 the U.S. Secretary of Energy. Moniz was pivotal in worldwide non-proliferation negotiations. He has been regularly in Brazil during the 1980's and 1990's, e.g., during meetings organized by the present authors. He contributed to numerous important theoretical works in Nuclear Physics \cite{Work1996,Work1998}.

The stories we tell in this brief review are divided into two main topics: the nuclear physics proper, and nuclear astrophysics. Everyone says that figures are worth a thousand words. But due to the limited space, we will only use words (and some equations, as we are physicists, OK?) to tell our stories and point to a few open questions in this intriguing, beautiful, and extremely difficult field of science. 

In Section II we present an account of the different facets of nuclei and discuss the major advances in nuclear structure and put special emphasis on the recent research activities in the area of neutron- and/or proton-rich nuclei, which have been produced and studied since the 80's. In Section III we discuss  reaction studies and discuss the different theories developed for the purpose. In Section IV we discuss the new nuclei we have discovered in the last few decades. In Section V we discuss the importance of nuclei in the Cosmos, and in the synthesis of elements. Finally, in Section VI,  concluding remarks are given.

\section{Different Facets of Nuclei}

\subsection{Quantum Mechanics, Nuclei, and the Origins of Life} 

Nuclear Physics is the driving science behind stellar evolution. It determines energy production, element abundance and has ultimately led to the knowledge development in other basic sciences such as Astronomy. Imagine a world without Astronomy and the benefits its observations have brought to such mundane things as the human sense of time, distance and motion, through navigation, and the discoveries of new worlds on Earth and in the skies. The symbiosis between nuclear physics and astronomical observations has skyrocketed in the last decades with the launch of astronomy dedicated satellites  and the development of new observation technology. In the end, Nuclear Physics is the crucial  piece of science to interpret the observations.  

Arguably, the field of {\it nuclear astrophysics} started with the monumental work of  Hans Bethe, Subrahmanyan Chandrasekhar, and other great pillars of science. A humorous event might have happened (nobody really knows) in which during the 1930's Hans Bethe told his girlfriend under a bright night sky watch that he was the only person on Earth who knew how stars shine. His girlfriend was apparently not impressed with his statement because girlfriends under a bright sky are more interested in something else.  What Bethe knew and possibly nobody else did was that one gains a lot of energy by nuclear fusion of 4 protons into a $^4$He ($\alpha$-particle) nucleus. Two protons turn into neutrons by positron emission (beta-decay process), releasing about  28 MeV ($\sim 4.5 \times 10^{-12}$ J) of energy per event. At the time Bethe proposed a mechanism by means of which a series of reactions involving carbon and nitrogen nuclei served as {\it catalyzers} for the production of the   $\alpha$-particle (the CN cycle). The  catalyzers are recycled, and the process is analogous to the work carried out by enzymes in biological processes. The original model proposed by Bethe was later extended to include oxygen nuclei and is now known as the {\it CNO cycle} \cite{Bet39}. Many decades passed by until we fully understood that Bethe's model is responsible for  energy generation in massive, hot stars, only.  For example, in contrast to Bethe's model our Sun catalyzes $\alpha$-particles by means of  proton-induced fusion reactions, also  known as the {\it pp-chain} \cite{Ad11}. 

The {\it tunnel effect} is responsible for the long time scale of stellar burning. Without tunneling there would be no stars as we know them. It is not a coincidence that all main physics related to the understanding of how stars shine occurred in parallel with advances in Nuclear Physics. The more we learn about Nuclear Physics, the more we learn about stars and the universe.  Sorry, {\it stellar modelers}, you have to wait for us! The first application of the tunnel effect  occurred in Nuclear Physics  to understand the process of $\alpha$-emission by heavy nuclei. For example, $^{210}$Po emits $\alpha$-particles that tunnel through a Coulomb barrier to get free outside the nucleus. This was explained by Gamow, Gurney and  Condon who calculated nuclear  half-lives by $\alpha$-decay (or $\alpha$-emission) using first principles of quantum mechanics available during late 1920's \cite{Gam28,GC28}. The inverse process of capture of $\alpha$-particles or any other charged particle (such as proton capture) is also explained as an inverse tunneling process through a repulsive Coulomb barrier.  

The immediate conclusion in this initial story-telling about nuclei and stars is that there would be no life in our Universe, or even no Universe as we know it (or the little we know about it!), without quantum mechanics ruling the physics of the nuclear world. The relevance of nuclear physics for stellar evolution is deeply rooted in processes generating energy  and forming heavier elements. But there is a {\it caveat}: {\it nuclear physics is arguably one of the most difficult research areas of all sciences}. This is because the nucleon-nucleon interaction is not so well known and also because nucleons are composite objects ruled by the physics of quarks and gluons, effectively manifested through other particles  which mediate the nuclear force (i.e., the {\it mesons}). Moreover, nucleons in nuclei are not so many to allow for simplifications traditionally available to understand the physics of bulk objects. In other words, nuclear structure and nuclear reactions are very difficult to handle using  the underlying theory of quantum mechanics. In more mundane words, {\it nuclear physics research is not for everyone.} Only the strong and stubborn scientists survive in nuclear physics research. We hope that this statement will only motivate graduate students to choose research in nuclear physics. That is because difficult problems are more challenging and motivating for real scientists. 

It is extremely likely that life would not exist without the presence of the elements we have identified in nature. The light elements were produced in nuclear reactions during the {\it Big Bang} and heavier ones after the formation of the first stars. Carbon is a crucial element for life, although it is not ruled out that other elements such as silicon might be the basis of life in other strange planetary environments \cite{Dav14}. A crucial step in understanding the origins of life as we know was the discovery of the {\it triple-alpha capture} process when three alpha-particles join to form a carbon nucleus. It was early concluded from many experiments that there would be no possibility to form carbon as abundant as we observe without resorting to  the triple-alpha process, which is part of a path to the formation of heavier elements. If the triple-alpha capture process would be non-resonant then the reaction rate would also be too small in a stellar environment. It was necessary to postulate the existence of a resonance  in the continuum of the beryllium nucleus (unbound) and  also in the carbon nucleus close its  alpha emission threshold, more precisely at 7.65 MeV  \cite{Hoy54}. The resonance in beryllium allows for a short time interval for the collision with another alpha-particle to occur, therefore the reason for coining it as the triple-alpha process. The resonance in carbon would be responsible for the large value of the capture cross section necessary to explain the existence of ``abundant" carbon atoms in the Universe.
 
This purely theoretical hypothesis, due to Sir Fred Hoyle \cite{Hoy54} was later confirmed by William Fowler and collaborators with an experiment at the Kellogg Radiation Laboratory at Caltech \cite{Fow84}. The resonance is known as the {\it Hoyle state} or the {\it nuclear state of life}.  Some  think of it as a support to the {\it anthropic principle}: we exist, therefore this state \cite{Fow84,Hjo11} in carbon is necessary, or has to be preconceived. Many scientist dislike the use of this principle in science and even argue that there is no principle at all, including the Nobel laureate Steven Wienberg \cite{SW87}. Such statements are {\it not even wrong}, as used to say another Nobel laureate Wolfgang Pauli. Believing in it is very much like an ornament fish's thinking: there must be a God, otherwise who would throw food in my aquarium? It would be embarrassing to find out that human brains are only as far able to think as a fish. Despite such spiritually inspired nonsense,  due to its immense relevance, the triple-alpha reaction still leads to genuine theoretical and experimental interest \cite{Fyn05,Hjo11,Mei15}.

\subsection{Gluons, Quarks and Nucleons} 

In 1949, again Hans Bethe \cite{Bet49} introduced the concept of {\it scattering length} and {\it effective range}  to describe nucleon-nucleon scattering at low energies. To simplify his description, in the absence of the Coulomb interaction the scattering phase shift expansion for a partial wave $l$ is given by
\begin{equation}
k^{2l+1} \cot \delta_l(E) = -{1\over a_l} + r_l {k^2\over 2} + \cdots \ , \label{bethe}
\end{equation}
where $E=(\hbar k)^2/2\mu$, $\mu$ is the reduced mass of the system, $a_l$ is the scattering length and $r_l$ the effective range. 

Bethe had no idea of the existence of an inner nucleon structure. Neither did Hideki Yukawa when he introduced a simple model to explain nuclear forces based on pion exchange \cite{Hi35}. His model was later extended to include other mesons such as the rho and omega mesons. Now that we know that nucleons are made of quarks and gluons, we also know that the {\it meson-exchange theory} of nuclear forces is an {\it effective theory} by means of which the quarks and gluons degrees of freedom are hidden. Such a meson-exchange approach is very much like Bethe's theory of Effective Range Expansion (ERE), Eq. \eqref{bethe}, which describes many features of the nucleon-nucleon forces in terms of  the scattering length and the effective range. They hide the nuisances of the nuclear interaction at very low energies.   

Since we know that quarks and gluons flourish within nucleons, we also discovered that  the QCD Lagrangian \cite{Gel95}
\begin{equation}
{\cal L}_{QCD} = \bar{\psi_n}\left[i(\gamma^\mu D_\mu)_{mn}-m_q\delta_{mn}\right]\psi_m-{1\over 4} G_{mn}^aG^{mn}_a\ , \label{QCD}
\end{equation}
is the basic theory describing the inner works of the nucleon-nucleon interaction.
In Eq. \eqref{QCD}, $\gamma^\mu$ are Dirac matrices, $\psi_n(x)$  is the quark field, depending on space and time, indexed by $m,\ n, \cdots$, and $m_q$ are quark masses.   Moreover,
\begin{equation}
G^a_{\mu \nu} = \partial_\mu \mathcal{A}^a_\nu - \partial_\nu \mathcal{A}^a_\mu + g f^{abc} \mathcal{A}^b_\mu \mathcal{A}^c_\nu \,,
\end{equation}
where $g$ is a coupling constant, $f^{abc}$ are the structure constants of SU(3) and $\mathcal{A}^a_\mu(x)$ are non-Abelian gluon fields, also  functions of space and time, in the adjoint representation of the SU(3) gauge group, indexed by $a, \ b, \cdots$.  The $G^a_{\mu \nu}$  are the QCD analogous to the electromagnetic field strength tensor, $F_{\mu\nu}$, of quantum electrodynamics \cite{PS95}.

There has been a strong effort to obtain nuclear properties from first principles by  solving Eq. \eqref{QCD}. The first step, namely, reproducing hadron ground state and resonances has been rather successfully achieved, especially for light hadrons, by means of, e.g., Lattice QCD (LQCD)   \cite{PDPG,BalS92,Lepa05,FH12,usqcd}.
The LQCD is formulated to solve Eq. \eqref{QCD} on a grid or lattice of points in space and time \cite{FH12}. There is still a long list of hadronic problems to be solved before a fully consistent set of results is achieved, also for heavy hadron masses. It more difficult to describe scattering states from LQCD \cite{Berk15} than hadronic ground state properties. It will certainly take decades or many human generations to extend the LQCD calculations to describe ground and excited state properties of many nucleon systems. First attempts in this direction are just at their infancy stage \cite{Sav15}. 

Using LQCD to obtain many bound and unbound nuclear properties also seems unnecessary and far fetched. It is much like {\it killing a fly with a cannon}. The world ``effective" is in the root of a better solution for the problem. One does  not need to use all the complex non-linear features of QCD for low energy nuclear physics because such features arise mostly at short ranges which are rather unaccessible in low energy processes. Thus, one needs an effective theory to handle nucleon-nucleon and multi-nucleon interactions in the same way as Bethe's ERE theory was developed for scattering theory with wave mechanics. In 1979 Steven Weinberg \cite{Wei79} proposed the use of {\it Effective Field Theories} (EFT) for low energy processes. In nuclear physics it means something like replacing the Lagrangian \eqref{QCD} by
\begin{equation}
{\cal L}_{QCD} \longrightarrow {\cal L}_{EFT} = N^\dagger \left( i\partial_t + {\nabla^2 \over 2m_N}\right) N + \left({\mu \over 2}\right)^{D-4} \left\{    -C_0  (N^\dagger N)^2 +{C_2\over 8} \left[ N^\dagger N \left( N^\dagger \overleftrightarrow{\nabla}^2 N\right)\right] \right\} + \cdots \ , \label{EFT}
\end{equation} 
where $N$ are nucleon fields for the isospin doublet, $N=(p,n)$, which are functions of spacetime, and $ \overleftrightarrow{\nabla}^2 =  \overleftarrow{\nabla}^2 + 2 \left(\overleftarrow{\nabla} \cdot  \overrightarrow{\nabla}\right)+ \overrightarrow{\nabla}^2$.   In this expression, $\mu$ is a constant to give the second term a correct dimension for any $D$. The constants $C_i$ are called {\it Low Energy Constants} (LEC) and carry the hidden information on the short-range physics, i.e., they carry the complex information about quarks and gluons inside the nucleons. The simpler Lagrangian \eqref{EFT} carries all required symmetries from the underlying QCD Lagrangian and can lead to more precise results as those dots on the right hand side are replaced by more terms in the expansion. Some say that what matters in EFT is {\it how  one treats those dots}.

In Quantum Field Theory (QFT), given a Lagrangian one can solve scattering problems by ``reading" the Feynman rules out of it. Then one follows the usual procedure of QFT with regularization of integrals, renormalization due to infinite sums, and all those seemingly boring stuff. It is a completely different approach than solving the Schr\"odinger equation with a nucleon-nucleon potential. Thus, the solution of nucleon-nucleon problems using Eq.  \eqref{EFT} can be done with increasing levels of approximation, namely, to leading order (LO), next-to-leading order (NLO), next-to-next-to-leading order (NNLO, or N$^2$LO), etc. Thus, the level of precision of the calculation depends on how far does one go with the approximations (the dots). And they are (in principle) under control. In contrast, there is no predictive path to improve the Schr\"odinger potential method because nobody knows how to theoretically improve a phenomenological potential. EFT in fact gives a precise prescription of how to improve calculations \cite{ORK94,OK92,Kolc94,ORK95,Lepa97,Kolc98,Bed98,BedU99,Barn15}. Applications of EFT to  solar nuclear reactions and to other stellar evolution in general have been   reported in numerous works (see, e.g.,   \cite{Rup00,BHK02,Rup11,Rup13}. For example,  calculations of radiative capture reactions using EFT and lattice gauge theory have been reported (see e.g., Refs. \cite{Rup11,Rup13}). Calculations of the triple-alpha reaction and the state of life based on {\it chiral-EFT}  have also been the recent focus in Nuclear Physics research  \cite{Lee09,Epe11,Epel13,Epel14,Lan14}. 

Nuclear interactions are also strongly modified by the medium and solving the nuclear many-body Schr\"odingier equation is a task that has taken more than 80 years and is still being developed. The theoretical understanding of nuclear properties based on  the QCD Lagrangian is a slowly growing computational effort, despite the advent of fast computers \cite{Du08}.  It might require another 80 years of theoretical efforts. But the use of QFT, EFT, QCD, and renormalization group techniques for low energy nuclear  physics  has been useful to unify all potential models used in nuclear structure calculations. For example, a particular interaction coined as  $V_{low-k}$ is shown to parameterize a high-order chiral effective field theory two-nucleon force. In the theory, a cutoff dependence can be used as a tool to assess the error in the truncation of nuclear forces to two-nucleon interactions and introduce low-momentum three-nucleon forces \cite{Nog00,Schw02,Bog02b,Bog03,BogPR03,Schw03,Schw04,Stec10,Roth11}.

It is interesting that the nuclear physics community almost forgot about Bethe's ERE which was quickly adopted in other fields. Nuclear physicists in the 1950-1990 were mostly interested either in nuclear spectroscopy in low energy nuclear physics, or in the study of a quark-gluon phase transition which might be discovered in relativistic heavy ion central collisions \cite{PBM07}. This phase transition might have occurred during the Big Bang and a colder version of it might also occur in the {\it core of neutron stars}. The research efforts and funding feeding the relativistic heavy ion community almost extinguished the  scientific interest in low energy nuclear physics. Concepts such as ERE were being long forgotten in nuclear physics, although thriving in other areas. However, in the last three decades or so, the ERE method resurrected  in the nuclear physics literature and spilled out to other fields. It was in fact another  nuclear physicist, Herman Feshbach from MIT, that developed the concept of {\it closed and open channels} in reaction theory and who also introduced the concept {\it Feshbach resonances}. Both of these concepts became cornerstone tools to study atomic  physics phenomena and in particular were well suited for the physics studied with ion and atom traps \cite{Fes58}. ERE is now a standard tool used to understand many features in {\it optical lattices} \cite{Bloc05}, {\it Bose-Einstein condensates} \cite{Bose24,Einst25}, or the {\it unitary Fermi gases} \cite{Aur07,Forb12,Bul11,Bul14}.
  
\subsection{The Nuclear Chart, Varieties of Elements}

Assuming that we fully understand the nuances of the nucleon-nucleon interactions and how it emerges from QCD, we still would like to know how the nearly 250 stable nuclei emerge from a consistent theory. Most of these nuclei are composed of even number of protons and even number of neutrons, namely, even-even nuclei (see, e.g., \cite{Bertu07}). But even-odd and odd-even nuclei are also abundant. On the other hand, the number of odd-odd stable nuclei can be counted in our hands, nominally: they are rare and amount to only 5. Among these, 4 are very light, $^2$H, $^6$Li, $^{10}$B, and $^{14}$N. The fifth stable odd-odd nucleus is very heavy and exists in an. excited. isomeric state, $^{180m}$Ta. It has a high spin, $9^{-}$, which makes its decay to the ground state, $1^{+}$, by $\gamma$-emission or through $\beta$-decay extremely unlikely since these decay processes only change the angular momentum by one unit. Thus to decay to the ground state, $^{180}$Ta, with spin and parity $1^{+}$, and whose half-life is about 8.15 h, there must be an exceedingly weak and thus very slow multi-photon emission or a similarly very slow $\beta$-decay. As such, this isotope of Tantalum is stable in this isomeric state ($E^{\star} = 77.1 keV$), and it is the only stable nucleus to exist in such a state. Another feature of $^{180m}$Ta is that it is the rarest isotope in nature amounting to only 0.012\% abundance in the natural ore composed mostly of the odd-even nucleus $^{181}$Ta (99.988\%) \cite{NNDC}. 

Isotopes of neutron-rich nuclei extend till the {\it neutron drip line}, where the neutron separation energy is zero, while isotopes of proton-rich nuclei are fewer and extend to the {\it proton drip-line} which lies closer to the valley of the stability than the neutron drip line. The drip lines act as natural boundaries of bound nuclei. One important fact to mention is the nonexistence of stable nuclei with A = 5 and 8. This fact has an important consequence on Big Bang nucleosynthesis and especially on the carbon production. Recently it has been possible to produce neutron-rich and proton-rich radioactive nuclei as secondary beams, which allowed researchers to extend the study of nuclear structure to these exotic species of nuclei \cite{BCH93,Bertulani02}. This area of research has become a priority in many countries and major projects involving the construction of large facilities of accelerators are in the process of being developed in Germany, the USA, China, Japan and  elsewhere. In Brazil, the activity in this area is centered in RIBRAS, installed in the Nuclear Physics Laboratory of the Institute of Physics of the University of S\~ao Paulo \cite{Lic03,LLG14}. 

Another important extension of nuclear research has been in the production of the so-called {\it superheavy elements}, with proton numbers exceeding 100. These elements are produced through the fusion of a medium mass nucleus with a heavy one resulting in an excited compound nucleus that, after emitting several $\gamma$-rays and cooling down, starts emitting $\alpha$-particles till reaching the superheavy element that survives enough time to be studied both physically and chemically (see, e.g.. \cite{Fus11}). Some of these elements are: Z = 101, Mendelivium Md, Z = 102, Nobelium, No, Z = 103 Lawrencium, Lr, Z = 104, Rutherfordium, Rf, Z = 105, Dubnium Db, Z = 106, Seaborgium, Sg, Z = 107, Bohrium, Bh, Z = 108, Hassium, Hs, 109, Meitnerium, Mt, Z = 110, Z = 112, etc. \cite{Sob66,Mye66,Hof00,Cwi05,Sob07,Oga07,Hof07,Oga11,Khu14}.

There is a huge effort in Nuclear Physics to understand how nuclear masses emerge from a consistent theory of the nuclear forces. For medium and heavy mass nuclei, the mean field theories are the tool of choice. 
Starting from a nucleon-nucleon interaction including two and three-body terms, such as
\begin{equation}
v=\sum_{i<j }v_{ij}^{(2)} + \sum_{i<j<k} v^{(3)}_{ijk} \ ,
\end{equation}
the energy or nuclear masses can be calculated by means of \cite{RS04}
\begin{equation}
E=\sum_i\left< i \left| {p^2\over 2m_N} \right| i\right> +{1\over 2} \sum_{i,j}\left< ij\left| v_{12} \right| ij\right>_{\cal A}+
{1\over 6} \sum_{i,j,k} \left< ijk\left| v_{123}\right| ijk\right>_{\cal A} \ ,
\end{equation}
where $m_N$ is the nucleon mass, and $\cal A$ means anti-symmetrization of the matrix element. If a Skyrme contact-like interaction is used for $v$, it has been shown \cite{BV72} that the above functional can be easily calculated if the nucleon wavefunctions $\left.|i\right>$ are obtained from an iterative procedure such as the Hartree-Fock method.  Medium corrections of the nucleon-nucleon interaction by means of the Brueckner method  or the inclusion of pairing by means of a BCS or the Bogoliubov method,  can improve calculations of the functional \cite{RS04}. Inclusion of relativistic effects have also shown to be of relevance \cite{Lal04,Nik11}. Several groups are pursuing these calculations, e.g., those in Refs. \cite{Schu15,Eri12,Ben03,Ber09,GC14,Du14,Cha15,Nik11,AB13}, and many open questions remain as, e.g. how well one can treat paring in nuclear matter.  The Beijing group has made many relevant contributions to the progress in this research topic \cite{Zha15,Lian15,Meng06a,Meng06,Lian08,Zhou03,Meng96,Meng98,Zhao11}. 

One of the most anxiously expected developments in this field is the use of time-dependent many-body techniques to unravel real-time excitation, decay, transfer and fragmentation of nuclei during nuclear collisions. It is a very hard task which requires the use of supercomputers. Advanced work on this subject has been reported in Refs. \cite{Sim10,Bul11b,Bul13,Kaz13,Stet15}.  

\subsection{Nuclear Sizes}
Owing to the great amount of energy required to excite or change nuclear structure, as compared to atoms, it is difficult to use external fields such as electric or magnetic fields to change the properties of nuclei or the basic nucleon-nucleon interaction. This is to be contrasted to the case of atoms, where the atom-atom interaction can be altered through the application of an appropriate magnetic field which affects the so-called Feshbach resonances. With this effect the atom-atom interaction can be changed from repulsive to attractive, and going through zero, making possible the study of cold gases under these different conditions. In nuclei no such liberty is available and the only way available for inflicting changes is the direct interaction with other nucleons (capture reactions) or other type of hadrons. By adding more neutrons or protons, new types of nuclei are produced as the case in nucleosynthesis. Some of these changes have become possible with the advent of the field of nuclear fragmentation in which nuclei are broken up and fragments rich in neutrons or protons are separated and further accelerated as secondary beams, whose properties are then discerned and analyzed.  Nuclei, such as $^{11}$Li, have been shown to have an rms radius as large as that of the heavy nucleus of lead, $^{208}$Pb! Thus, larger nuclei with special properties such as the existence of a {\it halo} of excess neutrons or protons (such as $^{11}$Be, $^{11}$Li, and $^{8}$B) can now be produced \cite{Tani96,SagH15}. 

One of the important features of nuclear structure is the fact that nucleons are identical fermions and are thus subject to the {\it Pauli exclusion principle}. Excess nucleons can only occupy the outer most orbits. Such restriction is removed if the added particles are other types of hadrons. One such example is the hyperon $\Lambda$ which is a strange nucleon whith a strangeness quantum number of 1. Once the $\Lambda$ is introduced into the nucleus, it does not suffer the effect of the exclusion principle and can migrate to the center of the nucleus \cite{Bau10}. Through the attraction it exerts on the other nucleons, these tend to aggregate closer to the center, making the normal nucleus shrink in size. In the {\it hyper nucleus} $^{7}_{\Lambda}$Li ( = $\Lambda$ + $^{6}$Li), the radius of $^6$Li is 20\% smaller than that in free space. It would be interesting to study such shrinkage in size in heavy nuclei, such as $^{208}$Pb, through the production and investigation of the superheavy hyper-nucleus $^{209}_{\Lambda}$Pb \cite{Wrob04}.

Other ways to influence the nuclear structure have been suggested. Very recently, multi-MeV zepto-second coherent laser pulse by backward Compton scattering of optical laser light on a sheet of relativistic electrons has been proposed to provide a huge number of MeV photons into the nucleus which could either create a plasma of nucleons or to excite collective modes. These lasers are being developed at ELI  \cite{ELI} and IZEST \cite{IZEST}. These developments in laser technology will enable nuclear researchers to go beyond the usual studies of the nuclear response, and to actually inflict changes in the properties of nuclei which are hitherto not possible and could reach the realm of freedom of manipulation practiced in cold atomic gases and Bose-Einstein condensation research.

\section{Nuclear Reactions}

\subsection{The Optical Model and Beyond}

Research in nuclear reactions accompanied closely the development of models of nuclear structure \cite{Ber13}. Two aspects of nuclear structure were in apparent conflict. The shell model which assumes a mean field felt by the nucleons and a corresponding long mean free path, and the Bohr model of the compound nucleus which asserts that the nucleons, once captured suffer many collisions with the other nucleons and with the mean field, which implies a short mean free path. These conflicting models were later shown to be different manifestation of the same underlying nucleon-nucleon force and the Pauli exclusion principle. A model for the nuclear reaction which is an extension of the shell model to positive energies, was then developed, where the nucleus is taken to exhibit refraction due to the real mean field and diffraction due to the absorption caused by the formation of the compound nucleus, was developed by Feshbach, Porter and Weisskopf \cite{FPW54}, and is based on the use of a complex average potential felt by the impinging nucleon. 

The {\it optical model} equation is
\begin{equation}
 (K + U)\big|\Psi^{(+)}\big> =  E \big|\Psi^{(+)}\big> \ .
\end{equation}
where K is the kinetic energy operator and $U({\bf r}) = V({\bf r}) - iW({\bf r})$ is the complex potential, whose real part, $V(\bf{r})$  is related to the mean field used in the shell model (at negative energies), while the imaginary part, $-W({\bf r})$, accounts for the flux lost into the formation of the compound nucleus which is treated separately using the statistical model. 
In fact, a simple manipulation of the above Schr\"odingier equation can be used to derive a continuity equation through which an absorption cross section can be derived,
\begin{equation}
\sigma_{abs}(E) = \frac{k}{E} \big<\Psi^{(+)}\big|W({\bf r})\big|\Psi^{(+)}\big> \ .
\end{equation}

The above cross section accounts for the compound nucleus, treated within the optical model, as a sink of flux. This cross section can be expanded into partial waves, yielding,
\begin{equation}
\sigma_{abs}(E) = \frac{\pi}{k^2}\sum_{l = 0}^{\infty} (2l + 1) T_{l}(E) \ , \label{Tl}
\end{equation}
where the $l$th transmission coefficient, $T_{l}(E)$ is 
\begin{equation}
T_{l}(E) = \frac{4k}{E} \int_0^{\infty} dr |\psi^{(+)}_{l}(r)|^2 W(r) \ ,
\end{equation}
and $\psi^{(+)}_{l}(r)$ is the radial wave function of the optical model with scattering boundary conditions. 

In the case of fusion cross sections, Eq \eqref{Tl} can be used with  $
\lambdabar
 =1/k = \sqrt{\hbar^2/2mE} = \hbar/mv$ being the reduced wavelength (remember the reduced Compton wave length $\lambdabar_{C} = \hbar/mc$). Eq. \eqref{Tl} can be interpreted as if the cross section is proportional to $\pi\lambdabar^2$,  the area
of the quantum wave. Classically, different parts of the wave have different impact parameters and different fusion probabilities, $T_l$. Large impact parameters correspond to large angular momenta, leading to the weight $(2l+1)$. For fusion, it is very useful to use the concept of {\it astrophysical S-factor}, such that
\begin{equation}
\sigma_F(E)= {1\over E}S(E)\exp\left[-2\pi \eta(E)\right], 
\end{equation}
where $\eta(E) = Z_1Z_2e^2/\hbar v$, with $v$ equal to the relative velocity. The exponential approximately accounts for the barrier transmission. While $\sigma(E)$ decreases rapidly as the energy decreases, the astrophysical S-factor remains rather flat \cite{Huss00}.

Empirical optical potentials used in the analysis of elastic scattering of nucleons and heavy ions have been developed. Pion-nucleus scattering was also extensively studied using semi-microscopic optical potential based on salient features of the pion-nucleon scattering {\it t-matrix} including the excitation of the $\Delta$-resonance. In the case of heavy-ions, we mention the recent development of the S\~{a}o Paulo Potential, based on the double folding of the densities of the two nuclei in conjunction with an effective N-N interaction including non-locality owing to Pauli exchange \cite{SPP1,SPP2}. This potential has been quite successful in accounting for the elastic scattering of both stable and radioactive nuclei, after the addition appropriate {\it polarization potentials} that simulate the effect of the couplings to strongly coupled channels  \cite{HBC85}, such as the breakup one in the latter case \cite{Gome06}. Its use in {\it coupled channels} calculations has been  proven to be quite successful in dealing with the scattering of nuclei from deformed targets, and in fusion reactions \cite{FH80}. 

\subsection{Compound, Pre-equilibrium and Direct Reactions}
Low energy reactions are dominated by the mechanism of the formation and subsequent decay of the {\it Compound Nucleus} (CN), which is the nucleus formed from the capture of the projectile by the target nucleus. This compound system is formed at a relatively high excitation energy and angular momentum. According to Bohr's hypothesis, the formation and decay of the CN are independent as the whole process is statistical and it takes time for the system to equilibrate after the capture process. Accordingly it is assumed that the cross section to go from channel $\alpha$ (the entrance channel) to a final decay channel $\beta$, is for a given value of the angular momentum,
\begin{equation}
\sigma_{\alpha,\beta} = \frac{\pi}{k^2}(2l + 1)\eta_{\alpha} \eta_{\beta} \ .
\end{equation}
Summing over $\beta$ gives the total absorption cross section for angular momentum $l$, $\sum_{\beta}\sigma_{\alpha,\beta}= (\pi/k^{2}) \eta_{\alpha}\sum_{\beta}\eta_{\beta}$ which can be further simplified by recognizing that the transmission from channel $\alpha$, $T_{\alpha} = k^2/\pi (2l + 1))\sigma_{\alpha}$, can be written as $T_{\alpha} = \eta_{\alpha}\sum_{\beta}\eta_{\beta}$. Thus $\sum_{\alpha}T_{\alpha} = (\sum_{\beta}\eta_{\beta})^2$. These manipulations leads to $\eta_{\alpha} = T_{\alpha}/\sqrt{\sum_{\beta}T_{\beta}}$, and the cross section becomes,
\begin{equation}
\sigma_{\alpha,\beta} = \frac{\pi}{k^2} (2l + 1) \frac{T_{\alpha}T_{\beta}}{\sum_{\gamma}T_{\gamma}} \ . \label{HFd}
\end{equation}
The above expression is the {\it Hauser-Feshbach} (HF) cross section \cite{HF}. The importance of the work of \cite{HF} which is an extension of the {\it Ewing-Weisskopf} theory \cite{WW40}, is the observation that the transmission coefficients $T_{\delta}$, where $\delta$   designates any decay channel of the compound nucleus, including the entrance channel, is directly related to the absorption cross section of the optical model \cite{WW40}. The case of compound elastic, $\alpha = \beta$, is distinct as there are obvious correlations between the entrance and exit channels, and the HF cross section above has to be multiplied by a factor of 2 \cite{KKM79}. The angular distribution of compound nucleus cross section is symmetric around $\theta = \pi$, and in fact is isotropic as expected of a statistical process. 

Higher energy data on particle spectra have indicated deviations from the pure compound nucleus evaporation form, indicating the operation of another mechanism of decay. This particle emission process is associated with the excitation of 2p-1h (2 particle - 1 hole), 3p-2h, etc, configurations in the compound nucleus before a fully equilibrated system is reached. Such pre-equilibrium emission has been the subject of investigation both experimentally and theoretically. The angular distribution of the emitted particles is, however, not isotropic, indicating the contribution of yet another mechanism besides that of the CN. This other mechanism involves the excitation of the target nucleus through the excitation of 1p-1h, 2p-2h configurations with the projectile particle remaining in the continuum. The combined effect of multistep compound and multistep direct processes constitutes the model of pre-equilibrium reactions which is used in Ref. \cite{FKK}. For a recent review of compound nuclear pre-equilibrium reactions see Ref. \cite{CEH}.

\subsection{Feshbach's Theory}
Nuclear reaction data indicate clearly the existence of two types of processes. The ones dominated by the compound nucleus which are slow processes and treatable with the statistical model of Bohr and formalized through the Hauser-Feshbach theory, and others, fast processes, describable by the optical model equation as generalized to many channels. Processes, such as inelastic scattering, transfer reactions, and breakup reactions are commonly called direct reactions and characterized by forward peaked angular distributions with  oscillations indicative of their coherent, non-statistical nature. To accommodate both types of reactions, the slow, compound ones and the fast, direct ones, Feshbach \cite{Feshbach58, Feshbach62}, developed a formal theory based on projection operators. 

Call the open channels space projection operator, $P$, and the closed channels space $Q$, with $P + Q$ = 1 and $PQ = QP$ = 0, $PP = P$, $QQ = Q$, then the Schr\"odingier equation of the colliding nuclear system can be written as,
\begin{eqnarray}
(E - PHP)P\big|\Psi\big> &= &PHQ \big|\Psi\big>\\
(E - QHQ)Q\big|\Psi\big> &= &QHP \big|\Psi\big> \ .
\end{eqnarray}

These equations summarize the whole theory of nuclear reactions. To describe the direct reactions one eliminates the compound nucleus $Q$-space and averages out the corresponding resonances, to obtain the following effective equation,
\begin{equation}
\left(E - PHP - PHQ \Big<\frac{1}{E - QHQ}\Big>QHP\right)P\big|\Psi\big> = 0 \ .
\end{equation}

The Hamiltonian $H$ in the above equations is the sum $H = K + h_1 + h_2 + V_{1,2}$, where $K$ is the kinetic energy operator, $h_i$ is the intrinsic hamiltonian of nucleus $i$, and $V_{1,2}$ is the real interaction operator between the two nuclei. The effective polarization operator 
$$PHQ \left<{1\over (E - QHQ)}\right>QHP = PHQ \left[{1\over (E - QHQ + iI)}\right]QHP \ ,$$ 
where the energy averaging interval, $I$, is much larger than the average width of a resonance. This equation
is complex and by adding it to $PV_{1,2}P$, defines the complex optical potential operator, 
\begin{equation}
V_{eff} = PV_{1,2}P + PHQ \frac{1}{E - QHQ + iI}QHP,
\end{equation} and the optical model equation becomes,
\begin{equation}
(E - K - h_1 - h_2 - V_{eff}) \overline{P|\Psi>} = 0 \label{CC} \ .
\end{equation}

This equation is an equation for the optical wave function $\overline{P|\Psi>}$ which contains many components (channels). As such, it is a set of coupled channels equations that describe the direct reactions involving the channels projected by $P$. The compound nucleus is now completely hidden in the complexity of $V_{eff}$. To extract the contribution of the compound nucleus, one needs to resort to statistical considerations involving the fluctuation component of $P|\Psi>$. The treatment of this issue relies on the use of assumed random properties of the $P-Q$ couplings and  the whole apparatus of quantum chaotic scattering theory is employed. The Hauser-Feshbach cross section and the Ericson correlation function \cite{Eric60,Eric63} are then obtained, making the above theory a complete one.

The set of equations Eq. \eqref{CC}, constitutes the Coupled Channels Theory (CCT) of nuclear reactions. Most reactions at low and intermediate energies require treatment with the CCT  \cite{Breit95}. At higher energies or weak couplings, a perturbative treatment is adequate, through the use of the Distorted Wave Born Approximation (DWBA). This theory has been very useful in the study of the nuclear structure as the amplitude is linear in the coupling that induces the transition and accordingly spectroscopic information are unambiguously extracted \cite{Sat83}. Variance of this theory, which is commonly used in the case of transfer of nucleons in cases involving deformed nuclei, is the Coupled Channel Born Approximation (CCBA), where the distorted waves are generated from a coupled channels calculation, whereas the transfer coupling is treated perturbatively. Such theory is still in use in the case of the scattering of exotic neutron- and proton-rich nuclei from deformed target nuclei \cite{Satchler-Love}.

\subsection{Coupled Channels}

The advent of radioactive beams has brought into focus several important features of reaction dynamics. The CCT is an example. The wave function $\overline{P|\Psi^{(+)}>}$ is taken to be composed of a sum of several terms representing the important channels operating in the collision of a stable, or neutron- or proton-rich projectile with a target nucleus. Writing $P= P_{el} + P_{bup}+ P_{trans} + P^{\prime}$ renders the optical model equation a set of coupled equations for the elastic channel, $P_{el}$, the breakup channel, $P_{bup}$, the transfer channel, $P_{trans}$, and the other channels that are treated in an average way, $P^{\prime}$. For exotic halo nuclei, such as the one-neutron halo $^{11}$Be, the two-neutron halo $^{11}$Li, $^{20}$C, or one-proton halo, $^{8}$B, the breakup coupling is quite important owing to the close to threshold breakup channel. As such, it becomes important to consider $P_{el}$ and $P_{bup}$ in the projected P-space optical equation, when treating elastic scattering. 

The adiabatic model, i.e., the neglect of the breakup Q-value, is generally used at higher energies, as done in \cite{Jim-Jef}, and \cite{Ron}. In \cite{Jim-Jef} the Glauber approximation while in \cite{Ron} the usual adiabatic Schr\"dinger model are used for the radial wave function. The result of the analysis of the total reaction cross section in \cite{Jim-Jef} of $^{11}$Li,  $^{11}$Be, and $^{8}$B and a $^{12}$C target, demonstrated that the breakup channel coupling necessarily increases the reaction cross section and accordingly requires a larger rms radius of the halo nucleus, as seemed to be required by the data on the interaction cross section (the total reaction cross section without the projectile inelastic or breakup contributions) at the time, which were analyzed using the optical, eikonal, approximation \cite{Tan85,Tan88}. This result is easily understood using the general structure of the transmission coefficient $T_{l}(x) \equiv T(b, x) = 1 - \exp{[-2\chi(b, x) ]}$, where $\chi(b, x)$ is the imaginary part of the eikonal phase. $b$ refers to the impact parameter and the parameter $x$ refers to the separation distance between the core and the center of mass of the excess nucleons. The original, Tanihata analysis \cite{Tan88}, considered the optical average $\exp{[-2\left<\chi(b, x)\right>_{x}]}$ whereas  the adiabatic model deals with $\left<\exp{[-2\chi(b, x)]}\right>_{x}$. Jensen's inequality \cite{AH09,JG00} comes into play to dictate that $\left <\exp{[-2\chi}]\right>_{x} \geq \exp{[-2\left<\chi\right>_{x}]}$, and thus the  transmission coefficient in the adiabatic model $T_{ad}(b) = 1 - \left <\exp{[-2\chi}]\right>_{x}$ is smaller than the optical transmission coefficient, $T_{opt}(b) = 1 - \exp{[-2\left<\chi\right>_{x}]}$ rendering the reaction cross section $\sigma_{R, ad} = 2\pi \int_{0}^{\infty} b db T_{ad}(b)$, smaller than the optical one, $\sigma_{R, opt} = 2\pi \int_{0}^{\infty} bdb T_{opt}(b)$. To compensate for the reduction in $\sigma_{R, ad}$ compared to $\sigma_{R, opt}$, a larger rms radius, e.g. for $^{11}$Li, $<r_{11}^2>$ = 3.55 fm,  than the one extracted by Tanihata \cite{Tan88}, 3.1 fm, is required to fit the data.The same considerations were applied in the treatment of the elastic scattering of $^{11}$Be + $^{12}$C \cite{Ron}. Improvement of the adiabatic model was then developed by the Pisa
\cite{BoBe01,Marg03,BBB04,Ibra05,Bla07,Kum11,Rav12,Fla12}, Commerce \cite{Bert03,Bert05}, Brussels \cite{Baye1,Baye2,Cape10}, Osaka \cite{OB09,OB10,Fuk14,Min14} and Seville \cite{Garc13,Vale14,Migu14} groups, through the relativistic {\it Continuum Discretized Coupled Channels} (CDCC) model and the {\it dynamical eikonal model}.

At low energies the fusion process becomes important. A lot of attention has been dedicated to the influence of the excess nucleons on the tunneling probability which dictates the value of the fusion cross section at near Coulomb barrier energies. The discussion of this process and other reaction processes at low energies requires the solution of coupled channels equations involving the projections $P_{el}$, $P_{bup}$, and $P_{transf}$ exactly. The breakup channel involves three or four clusters in the continuum, requiring for its treatment a formidable three- or four-body scattering calculation. In practice, however, the continuum is discretized into a finite number of pseudo-states treated as inelastic channels. The resulting CDCC equations, \cite{CDCC1,CDCC2,CDCC3,CDCC4}, are solved for a truncated number of pseudo states using known numerical methods. Recent improvement and extensions of the CDCC have been made \cite{descouvemont13,descouvemont15}. Computationally the CDCC relies on an ad hoc truncation of the discretized continuum by keeping only a few pseudo-states. The effect of the neglected pseudo-states on the convergence of the results is seldom analyzed. Recently, the idea of using statistical methods to treat these neglected pseudo-states was advanced \cite{BDH14}, and a model for a {\it Statistical CDCC} (sCDCC), based on Quantum Chaotic Scattering Theory (see below) is currently being developed \cite{BCDFH15}.

The results of CDCC calculations for fusion reactions have taught us several things. The excess nucleons in the halo nuclei results in two important effects. The excitation of low lying dipole mode, the {\it Pigmy Dipole Resonance} (PDR) (also known as the Ikeda resonance) \cite{Ik88,Auma05,Ber07c,ANa13,SAZ13,SPF13}, and the close-to-threshold breakup.
These two effects were found to influence the fusion in two distinct ways. The breakup coupling results in a suppression of fusion at energies above the top of the Coulomb barrier, whereas the dipole excitation results in an enhancement of fusion at sub-barrier energies \cite{Hus95,Gom12}. This latter effect is intimately related to the  longer extension of the matter density of the halo nucleus. An extensive effort, both experimental and theoretical, has been dedicated to the low energy reactions of exotic nuclei and several reviews were published, \cite{Canto06,HT12,Esbensen,Canto15}. 

The interest in fusion of neutron-rich nuclei stems from the desire to produce a compound system that could survive long enough to be studied and analyzed. Heavier nuclei with large amount of excess neutrons have been produced as secondary beams, and the existence of the Pigmy resonance was clearly established in, e.g., $^{134}$Sn \cite{Auma05,SAZ13,ANa13,SPF13}. These nuclei with a thick neutron skin, were produced at high energies. It would be very interesting to produce the neutron skin nuclei at lower energies and study their fusion with a heavy target to test the idea of an enhanced fusion and a potential production of a superheavy nucleus with a Z and A beyond the known ones. Such investigation will become possible in the near future with the new nuclear facilities being built around the world. 

It is worthwhile mentioning that recent experiments have also investigated the nuclear neutron skin by means of completely different physics methods. For example, we cite the $^{208}$Pb Radius Experiment (PREX) based on parity violation in electron scattering \cite{Abra12,Hor01}. Such experiments seem to be less dependent on the data interpretation base on a model description for scattering by pions \cite{Garc92}, protons \cite{Ray78,Star94,Clar03}, or antiprotons \cite{Trzc01,Lens09}.

\subsection{Incomplete Fusion}

Besides complete fusion \cite{HCD03}, which involves the capture of the whole projectile, the breakup may lead to the capture of one of the fragments, resulting in what is known as {\it incomplete fusion}. Other names of this process are used in the literature, such as massive transfer, inclusive non-elastic breakup, surrogate reaction, etc. This process is important at it supplies a mean to study the fusion of a nucleus  which is otherwise difficult to produce as a beam, such as neutrons (in a deuteron induced reaction). Several methods and theories have been proposed to calculate the incomplete fusion reactions. We mention the ones which are currently in use \cite{IAV,HM1,HM2,UT,ULT,HFM}. 

The standard treatment rely on the spectator model, which says that the two-cluster projectile $a = x + b$ interacts with the target only through the participant nucleus $x$, leaving the spectator $b$ unaffected and only suffer elastic scattering. The cross section for observing $b$ is invariably derived to be,
\begin{equation}
\frac{d^{2}\sigma}{d\Omega_{b}dE_{b}} = \frac{2}{\hbar v_a} \rho_{b}(E_b) \left<\hat{\rho}_x\big|W_{x}\big|\hat{\rho}_x\right> \ ,
\end{equation}
where $E_b$ is the energy of the outgoing fragment $b$, and $\rho_{b}(E_b) = m_{b}k_{b}/(8\pi^{3}\hbar^{2})$, is the density of states of $b$. The source function, $\hat{\rho}_{x}({\bf r}_x)$, is the wave function of the participant fragment $x$ in the projectile as it reaches the target nucleus. The imaginary part of the x-A optical potential is designated by $W_{x}({\bf r}_{x})$. 

In Refs. \cite{HM1,HM2,HFM}, the source function is the overlap 
\begin{equation}
\hat{\rho}_{x}({\bf r}_x) = \Big(\chi^{(-)}_{b}({\bf r}_b)\Big|\chi^{(+)}_{a}({\bf r}_x, {\bf r}_b) \phi_{a}({\bf r}_{b} -{\bf r}_{x})\Big>,\end{equation}
where $\chi$ is the optical model wave function (distorted wave), and $\phi_a$ is the intrinsic wave function of the projectile. The coupling interaction in Refs. \cite{IAV,HM1,HM2} was taken to be $V_{x,b}$ (post), while in \cite{UT,ULT}, the interaction is the difference in the optical potentials, $U_{x}({\bf r}_x)+ U_{b}({\bf r}_b) - U_{a}({\bf r}_a)$ (prior). Further, in Refs.  \cite{IAV,UT,ULT}, the incident wave function of $a$ is taken to be beyond the distorted wave. A debate has been going on concerning which of the above approaches is more appropriate for the calculation of incomplete fusion \cite{HM1,HM2,Ichimura,Surrogate}. This debate continues. 

The important point to mention here is that the structure of the cross section above is similar to the one suggested by Serber \cite{Serb47,But50}, namely the spectrum of the spectator fragment is proportional to the squared Fourier transform of the intrinsic wave function of the projectile, times the total reaction cross section of the participant fragment. Further  analysis was performed in Refs. \cite{HM1,HM2} using the eikonal approximation for the distorted waves gave for the incomplete fusion cross section the simple form of an integral over impact parameter of terms of the type, $\left<(1 - T_b)T_x\right>$, where $T$ is the optical transmission coefficient, and the average is over the internal motion of $b$ inside the projectile. If the $T_b$ is set to zero and the average is ignored one recovers the Serber expression \cite{Serb47},
\begin{equation}
\frac{d^{2}\sigma}{d\Omega_{b}dE_{b}} \sim \rho_{b}(E_b) |\phi_{a}({\bf k}_b)|^{2} \sigma_{x}(E_x) \ .
\end{equation}

The precise determination of the incomplete fusion is important not only for the purpose of capture reactions of only a part of the projectile, such as neutrons in deuteron induced reactions, but also in fusion studies as in many instances the data give the total fusion which is the sum of the complete fusion plus the incomplete fusion. Therefore to get the complete fusion one needs to subtract from the data a believable incomplete fusion cross section. The research in the area of incomplete fusion is currently pursued by several groups. A semiclassical model has recently been developed \cite{Marta}. In fact only in this year three papers have been written  
and will appear soon published in the journals \cite{Lei15,Potel15,Carlson15}.

\subsection{Quantum Chaotic Scattering}

A subject of continuos interest in nuclear physics research is the understanding of fluctuations in the cross sections. Back in the early 50's Wigner \cite{Wigner, Wigner2} recognized that it is meaningless to try to analyze the many resonances seen in the neutron capture cross sections in the region of overlapping resonances. He suggested that a statistical treatment is more appropriate and introduced the concept of random matrices for the purpose. The idea is to assume that the nuclear Hamiltonian be considered a member of an ensemble of random matrices and averages of different quantities be perfumed through  ensemble averages. The {\it Random Matrix Theory} (RMT) of nuclear reactions was born. The RMT was then greatly developed by Dyson \cite{Dyson} and application to fluctuation phenomena in other fields besides nuclear physics ensued. 

At the same time Ericson \cite{Eric60,Eric63,BS1963} introduced the {\it cross section correlation} function as a measure of the degree of coherence in the otherwise chaotic nuclear system. Through an analysis of the correlation function or the average density of maxima \cite{BS1963} one is able to extract the correlation width, which had been previously estimated by Weisskopf to be 
\begin{equation}
\Gamma_{corr} = {\overline{D}\over (2\pi)}\sum_{\beta} T_{\beta}, 
\end{equation} 
where $\overline{D}$ is the average spacing between the overlapping resonances. These developments were quite important as they are universal in nature and can be employed in many systems exhibiting fluctuations in the observables. This synergy, supplied by research in nuclear physics, attests to the richness of the ideas and concepts introduced in the field \cite{Ram12}. As a matter fact, recent application of RMT has been mostly in mesoscopic systems, such as electronic conductance in quantum dots and graphene \cite{HuPa00}. The test of RMT has been made possible using, among others, microwave resonators \cite{Weiden1,Weiden2}. 

The {\it Quantum Chaotic Scattering Theory} relies basically on an expression of the $S$-matrix which exhibits its relation to the random Hamiltonian of the system. It is normally written as,
\begin{equation}
S (\varepsilon)= \mbox{$\openone$} - 2\pi i W^\dagger (\varepsilon - H + 
i \pi W W^\dagger)^{-1} W \; .
\end{equation}

In the equation above $H$ is the random Hamiltonian which pertains to one of the universal classes of random matrix ensembles, the {\it Gaussian Orthogonal Ensemble}, GOE, the {\it Gaussian Unitary Ensemble}, GUE or the {\it Gaussian Simpletic Ensemble}, GSU, considered by Dyson and excellently reviewed by Bohigas and Giannoni \cite{Dyson,BG84}. The coupling matrix $W$ couples the internal degrees of freedom to the open channels, and is taken to be fixed (not random). The observables such as the average cross section and the correlation function are then calculated by performing an average over the ensemble to which $H$ pertains of products of $S$-matrices of the type given above. For more details we refer the reader to \cite{Weiden1,Weiden2}. 

Experimentally, the emergence of chaos in quantum systems can be verified by looking at the statistical properties and eigenvalues of the wave functions, and in in the fluctuation properties of the scattering matrix elements of some systems, such as quantum dots, quantum wires and Dirac quantum dots (quantum dots on a graphene flake, where the electrons are massless and obey the Dirac equation rather than the Schr\"dinger equation), etc.  Microwave billiards provide very useful systems to study quantum chaos,  because  they already contain a degree of chaoticity in their classical dynamics. In fact, the eigenvalues and wave functions of  quantum microwave billiards have been studied by Achim Richter's group \cite{DiAc15,Dietz14,Kum13}  with a high precision insight into quantum chaos phenomena.

\section{New Nuclei}

\subsection{Halo Nuclei, Efimov States, and Borromean Nuclei} 

A rather simple but ingenious experiment reported in 1985 by Isao Tanihata and his group on interaction cross sections of light nuclei close to the drip line was the seed of a new era in nuclear physics \cite{Tan85}. This experiment and others following it, have shown that some nuclei such as $^{11}$Li possess a long tail neutron distribution. The long tail is due to the low binding energy of the valence nucleons. It is simple feature, but one that nobody saw before, and it took Tanihata's genius to realize the apparently obvious nuclear property. His experiments were put within a nice context by Hansen and Jonson \cite{HJ87} who relied on the analysis of Coulomb breakup of $^{11}$Li, shown to be enormous due to the loosely bound character of the nucleus \cite{BeBau88}. After Hansen and Jonson's paper was published the community suddenly realized Tanihata's discovery and the number of works and citations to the emerging field of exotic, halo, and neutron-rich nuclei skyrocketed. 
 
These developments lead to the then existing nuclear facilities in Europe (GSI/Germany and GANIL/France), in the USA (NSCL/MSU), and in Japan (RIKEN) to almost entirely turn their beams and detectors to the production of radioactive beams. Experiments with {\it nuclei far from the stability} became  a routine in nuclear physics. The opening of this new research field in nuclear physics was one of the reasons why the  U.S. National Science Foundation decided not to close the NSCL facility in Michigan State University (MSU) in 1992.  The Indiana Cyclotron facility was not so lucky at the time, perhaps because they missed those historical developments.  Since then new facilities have evolved and planned in all continents, in particular we mention the new RIKEN radioactive beam facility and the  FRIB/MSU and FAIR/Darmstadt facilities presently under construction \cite{fairf,frib}.  In particular,  it took the relentless and diligent work of Konrad Gelbke and collaborators to elaborate and get funds for the FRIB/MSU facility in the U.S. Also in China and Korea other facilities are being planned or under construction. For a review of the theoretical efforts conducted at the period of 1990's, we suggest the reader to read Refs. \cite{BCH93,Zhu93,wong,Bertulani02}.

 With so much of science at stake it was imperative to obtain funding for new nuclear physics laboratories with the sole goal of studying nuclei far from the stability line.  It is worthwhile mentioning that before this era, most physics studied in those facilities had to do with central collisions  with the purpose to study the {\it equation of state} (EOS) of nuclear matter at high densities and temperatures \cite{Li08}. This EOS (pressure versus density), specially at low temperatures, is crucial for  understanding the physics of supernova explosions and neutron stars. In fact, masses and radii of neutron stars are constrained by the EOS of nuclear matter.  In contrast to the study of EOS,  the physics of radioactive secondary beams was mostly driven by peripheral, direct reactions. Stripping, Coulomb and nuclear excitation, two- and three-body breakup, and other direct reactions were and remain the principal tools to access spectroscopic information of interest to model and develop theories for short-lived nuclei, far from the stability line.  Their relevance for astrophysics purposes is discussed below.

Typically, neutron binding energies are of the order of 8 MeV for most stable nuclei. But neutron separation energies as low as  few keV have been observed in short-lived nuclei. The long (visible only in a logarithmic scale!) tail of the weakly-bound neutrons and their corresponding densities is known as ``halo" and the nuclei with this property are known as {\it halo nuclei}. Not only the small separation energies, but also nucleon-nucleon correlations are vital for the halo properties. As a typical example, neither $^{10}$Li nor two neutrons form a bound system, but together in $^{11}$Li they bind by about $300 $ keV \cite{NNDC}. The nucleus $^{11}$Li became a sort of nuclear superstar in the 1990's. There was not a single month or weeks without a nuclear physics preprint reporting either an experiment or a theoretical calculation on $^{11}$Li. {\it Nuclear physicists also love fashion, fame, and applauses}, not differently than anybody else.  

Three-body systems theorists were very happy to see that $^{11}$Li fell in their category of delicate structures arising from loosely bound three-body systems. In fact, $^{11}$Li is a prototype of a structure known as the {\it Efimov effect} \cite{Efi70}, i.e., the appearance of (many) bound states in a system of two-body subsystems with large scattering lengths  \cite{Efi70,Yam06}. Jan Vaagen named such systems as {\it Borromean systems}, because they reminded him of the heraldic symbol of the aristocratic Borromeo family from the fifteen century \cite{Zhu93}. In fact, the Borromean rings shown in the heraldic symbols are  three inter-connected rings, such that if one is cut loose, the remaining two also become free. 
A large activity in the area of three and few-body physics have profited from this early work on nuclear physics in the 1980's and 1990's.  Efimov states in Borromean systems  have become a common feature in atomic \cite{Kra06} and nuclear \cite{BH07} physics.

\subsection{New Magic Numbers, Clusters, Majorana Particles} 

In 1949 Maria Goeppert-Mayer and Hans  Jensen proposed that nucleon forces should include  a spin-orbit interaction able to reproduce the nuclear magic numbers 2, 8, 20, 28, 50, 82, 126, clearly visible from the systematic study of nuclear masses  \cite{Ma49,Ha49}.  Now we know that as nuclei move away from the line of stability ($Z\sim N$), where $Z$ ($N$) is the proton (neutron) number) those magic numbers might change due to correlations and details of the nucleon-nucleon interaction, e.g., the {\it tensor-force}  \cite{Ste13,Ots05,Suz03,Hon02,Otsu10,Yua12,Suz13}. These  and other properties of nuclei far from the stability and in particular those close to the {\it drip-line} (i.e., when adding one more nucleon leads to an unbound nucleus) increased enormously the interest for low-energy nuclear physics, in particular for the  prospects of its application to other areas of science \cite{Cot10}. The applications in astrophysics are evident. For example, the {\it rapid neutron capture process} (rp-process) involves nuclei far from the stability valley, many of which are poorly, or completely, unknown. 

It is not easy to predict what one can do with short-lived nuclei on Earth. Sometimes, they have lifetimes of milliseconds, or less. It is tricky to explain to the laymen the importance of such discoveries in the recent history of nuclear physics. For example, in 1994, the unstable doubly-magic nucleus $^{100}$Sn was discovered at the Gesellschaft f\"ur Schwerionenforschung (GSI), Darmstadt, in Germany \cite{Sch94}. At the same time the laboratory developed a new {\it cancer therapy} facility based on the stopping of high energy protons in cancer cells. To celebrate the new facility, the laboratory invited high-ranked authorities from the German government. They surely brought along the main news media in the country. Wisely, the laboratory management included the discovery of $^{100}$Sn alongside the proton therapy facility in the celebration agenda. At one moment a reporter asked a nuclear physicist what is $^{100}$Sn good for practical purposes, what was followed by a long silence  and a curious answer: ``I think one can develop better, lighter, cans." Maybe so, but such a can would last fractions of second! The only sure prediction for applications of short-lived nuclei is that it is impossible to understand stellar evolution without a dedicated study of their nuclear properties.

{\it Clustering phenomena} in nuclei is a difficult problem, which requires physics beyond the shell model (SM). The SM describes  protons and neutrons moving in shells. Other models start from a  molecular viewpoint that employs composite particles, such as the $\alpha$-particles as the building blocks for nuclear structure. A major task is underway to connect the two seemingly disjoint points of view and how nuclear excitations can emerge from a unified picture of the nucleus \cite{Iked68,Feld98,Itaga00,Free01,DeBa04,Beta02,Neff03,Okolo03,Voly06,Mic07,Mas07,Itaga07,Ito08,Roth10,Oko13,Free12}.  

Majorana had already shown in 1937 that if the neutrino would be its own anti-particle then the beta-decay theory would not change. If the neutrinos are their own anti-particle ({\it Majorana neutrinos}), then neutrinoless beta-decay in double beta-decay is also possible, with two neutrons being converted into two protons in the decay.  But this is a very rare process and extremely suppressed, what makes it difficult to access experimentally. There has been some early hints that the process might occur in nature  but up to the present date there has been no definitive proof of its existence \cite{Beri12,Gar14}. The shell-model and nuclear structure calculations are crucial to compare theory to experiments. The plethora of nuclear structure techniques involve, among others,  the interacting shell model \cite{Cau08,Men09}, the quasiparticle random-phase approximation \cite{Rod06,Kort07}, the interacting boson model \cite{Bare13} and the energy density functional method \cite{Pine10,Vaq13}.

\section{Nuclear Astrophysics}

\subsubsection{ \bf The Repulsive Coulomb}

Although stars are very hot, the kinetic energy of particles in their plasma are very small compared to the Coulomb repulsion energies, or the Coulomb barriers, in nuclei. For example, in the Sun's core the temperature translates to about 10 keV of relative energy of the nuclei. It is very difficult to measure reactions at such energies because of the smallness of the cross sections due to the {\it Coulomb repulsion}. In general, i.e., either for the Sun or for other stellar scenarios, many of the reactions of interest are not known at the level of accuracy required by stellar modelers, or they are simply not measurable directly at all. Many reactions also involve short-lived nuclei which are difficult, if not impossible, to manipulate.

One option for experimental nuclear astrophysicists is to use, e.g., stable or radioactive beams and study reactions with inverse kinematics techniques. Reactions with unstable nuclei use a stable target (e.g., a proton gas target) and by inference one can access information on cross sections of astrophysical interest.  This information relies heavily on reaction theory \cite{Ber13}. We cite a few  phenomena (only a few, indeed) of current interest using these techniques.

\subsection{Collective Resonances, Neutron Skins and Neutron Stars} 

At low energies, of a few MeV/nuclei, heavy nuclei are not able to surpass the Coulomb barrier and nuclei basically interact only via the Coulomb force. At high energies, above the Coulomb barrier, experimental techniques have been developed to extract Coulomb excitation events from other sort of concurring reactions. A plethora of processes are studied using this technique which has the advantage of using a well-known interaction and therefore the reaction mechanism is very much under control. A review of the applications of Coulomb excitation to several reactions of astrophysical interest is found in Refs. \cite{BB88,BG10,Tribb14}. One of the processes of interest for astrophysics is the excitation of  collective resonances above the particle emission threshold. The most notable of these are the so-called {\it giant resonances}. They exist in all sorts of modes. There are monopole, dipole, quadrupole and octupole vibrations, related to the character of nuclear shape vibration. They can also be of isoscalar or isovector character, due to their isospin vibration type ($T=0$ ot $T=1$). There are also electric and magnetic collective excitations.   

Collective vibrations in nuclei are intimately related to the compressibility of nuclear matter, or {\it equation of state} of nuclear matter. This equation give the density dependence  of the energy density in nuclear matter as
\begin{equation}
E(\rho,\alpha)=E(\rho,0)+S(\alpha)\alpha^2 +{\cal O}(\alpha^2),  \ \ \ \ \ \ \ \ {\rm   where} \ \ \ \ \alpha = {\rho_n - \rho_p \over \rho} \ , \label{eos}
\end{equation}  
with $\rho = \rho_p + \rho_n$ equal to the nuclear density. The energy $\epsilon(\rho)=E(\rho,0)$ is the energy for symmetric nuclear matter. From it one can calculate quantities of interest for astrophysics, such as the pressure, $P(\rho)=\rho^2 \partial \epsilon(\rho)/\partial  \rho$, and the {\it matter incompressibility}, $K=9\partial P(\rho)/\partial \rho$.  These quantities, together with the entropy density, are key to determine the relation between mass and radii of neutrons stars.

The EOS of nuclear matter can only be inferred by a intimate combination of theory and experiment because nothing on Earth can resemble the matter at extreme densities inside a neutron star \cite{BP02}. The main part of the EOS, i.e., the dependence of nuclear pressure on nuclear density can only be inferred in a small range of interest for neutron star modelers. These are often strongly guided by theoretical mean field calculations. Traditionally, theorists are able to calculate  the nuclear matter energy incompressibility modulus, $K$, using energy density functional theories, which start from the nucleon-nucleon interaction in the nuclear medium and calculates $E(\rho,\alpha)$ using a microscopic theory such as the Hartree-Fock, Relativistic Mean Field Theories, Random Phase Approximation, or variations of these theories \cite{Ben03,Ber09,GC14,Du14,Cha15,Nik11,AB13,Mas15}.    

It is well-known that although some effective mean field interactions can reproduce many nuclear properties quite well, their prediction strongly deviate from each other as one departs from the nuclear density saturation ($\rho=\rho_0 \approx 0.17$ nucleons/fm$^3$ )\cite{Brow00}.
As an example of such microscopic calculations, by comparing Hartree-Fock-Bogoliubov results for the excitation of giant monopole resonances in nuclei one can infer what sort of density functionals are most appropriate to reproduce the experimental data. Giant monopole resonances are the most direct compressible mode of a nucleus and and clearly related to the compressibility modulus.  Inspecting the most accepted forms ot the microscopic interactions has shown that a compressibility modulus, $K\simeq 200$ MeV \cite{AB13} can be obtained, although values of  $K\simeq 230$ MeV are also acceptable by studies of additional mean-field calculations for the excitation and decay of other giant resonance modes. 

Relativistic heavy ion collisions also induce a matter compressibility during a short-time interval, with consequences such as side-flow of nuclear matter which can be measured and compared to microscopic calculations usually done with transport theories such as the {\it Boltzmann-Uehling-Uhlenbeck} (BUU) equation \cite{Dan02,Li08},
\begin{align}
  \frac{\partial f}{\partial t}+\left(
\frac{\mathbf{p}}{m_N}+\mathbf{\nabla }_{\mathbf{p}}U\right)
\cdot\mathbf{\nabla}_{\mathbf{r}}f-\mathbf{\nabla
}_{\mathbf{r}}U\cdot\mathbf{\nabla}_{\mathbf{r}}f&=
  \int d^{3}p_{2}\int d\Omega\;\sigma_{NN}\left(  \Omega\right)
\left\vert
\mathbf{v}_{1}-\mathbf{v}_{2}\right\vert \nonumber \\
&\times\left\{  f_{1}^{\prime}f_{2}^{\prime}\left[  1-f_{1}\right]
\left[ 1-f_{2}\right]  -f_{1}f_{2}\left[  1-f_{1}^{\prime}\right]
\left[
1-f_{2}^{\prime}\right]  \right\}  . \label{BE}%
\end{align}
Here, the number of particles in the volume element $d^{3}rd^3p$ at time
$t$, is given by  $f\left(  \mathbf{r,p},t\right)  d^{3}rd^{3}p \label{distbf1}%
$, in terms of the distribution
function $f\left(  \mathbf{r,p},t\right)$. The \textit{mean-field}, $U\left(  \mathbf{r,p}%
,t\right)$ accounts for the effect
of each particle interacting with all others, with forces on each particle given by $-\nabla_{\mathbf{r}}U\left(
\mathbf{r,p},t\right)$ and
$-\nabla_{\mathbf{p}}U\left( \mathbf{r,p},t\right) .$
The distribution function
changes due to nucleons leaving (or
entering) the volume $d^3rd^3p$,  accounted for by the {\it collision
term} on the right side, where $\sigma_{NN}$
is the nucleon-nucleon scattering cross section, ${\bf
v}_1$ and ${\bf v}_2$ are the velocities of two colliding nucleons, and $\Omega$ their scattering angle.
The factors $(1-f)$ account for Pauli blocking of final occupied states.

In further constrain the EOS of nuclear matter, the function $S$ in Eq. \eqref{eos}, called by the {\it symmetry energy} is needed. It is usually expanded around $x=0$ where  $x=(\rho-\rho_0)/3\rho_0$, yielding
\begin{equation}
S={1\over 2} \left. {\partial^2 E\over \partial \alpha^2}\right|_{\alpha=0}=J +Lx+{1\over 2} K_{sym}x^2 +\cdots
\end{equation} 
where $J$ is the bulk symmetry energy, $L$ determines the slope, and $K_{sym}$ the curvature of the symmetry energy at saturation, $\rho=\rho_0$  ($\approx 0.17$ nucleons/fm$^3$). The use of heavy ion central collisions and the interpretation of experimental data using transport equations has been proven to be a good method to study the effects of symmetry energy in nuclei and in nuclear matter (see, e.g., Refs. \cite{BaLi00,BaLi02,Chen05}).

Collective vibrations of nuclei are also a good tool to investigate the symmetry energy. Some features of neutron rich nuclei are also though to be very sensitive to the the symmetry energy, such as their neutron skin $\Delta r =r_p-r_n$, where $r_{p(n)}$ is the proton (neutron) matter distribution \cite{Ber07a,Cen10}. Collective vibrations in neutron-rich nuclei at low energies are usually so-called {\it Pigmy Resonances} or {\it Ikeda resonances} \cite{Ik88,SAZ13,Wie09,Ros13,SAZ11}. According to equation (14) of Ref. \cite{Ber07b}, the energy of pigmy resonances, $E_{PR}$, are elated to the neutron skin, $\Delta r$, and the neutron excess, $N_{exc}$, by means of
\begin{equation}
E_{PR} = \left( {3S_n \hbar^2 \over 2 \Delta r R m_N N_{exc}}\right)^{1/2} \ , \label{skin}
\end{equation}
where $R$ is the nuclear radius and $S_n$ the neutron separation energy. Inserting typical values for $S_n$, $R$, $N_{exc}$ and $\Delta r$ for medium mass neutron-rich nuclei, one obtains $E_{PR} = 5 - 8$ MeV. 

Study of pigmy resonances helps constrain the symmetry energy dependence of the EOS of nuclear matter in neutron stars \cite{Bru68,Li08}. They are also related to the neutron skins in nuclei, as shown in Eq. \eqref{skin}.  \cite{TB01,Furn02,Ber07a}. Pigmy resonances are a possible energy sink during supernovae explosions. 
Their existence can lead to noticeable  changes in the abundance of heavy elements by means of rapid neutron capture reactions, or {\it r-processes} for short \cite{Go98}.  In the process a neutron capture occurs on a very short time-scale, $\sim 0.01 - 10$ s. This is often much faster than accompanying $\beta$-decay processes occurring between the neutron captures. The r-process path therefore approaches the neutron drip-line as nuclei get more neutron-rich. A large number of isotopes in the range $70 < A < 209$ are produced in this way. The stellar site of r-processes are not well known. It might occur during supernovae explosions, or during neutron star mergers processes. 

The nuclear dipole polarizability  $\alpha_D$ has recently been suggested \cite{BHR03,RN10,Piek11} as an alternative observable
constraining both neutron skin and symmetry energy. The polarizability
is related to the photoabsorption cross section
\begin{equation}
\alpha_D={\hbar c\over 2\pi^2e^2} \int {\sigma_{abs}(\omega) \over \omega^2} d\omega ,
\end{equation}
where $\omega$ is the photon energy. Because of the inverse energy weighting, the low energy response sensitive to the symmetry energy is probed. Recently, experiments carried out in Japan and led by Peter von Neumann-Cosel allowed to infer the dipole polarizability of several nuclei  \cite{Tam11,Pol12,Krum15}. Early work by Takashi Nakamura and Thomas Aumann \cite{Auma05,SAZ13,ANa13,Wie09,Ros13} have also unravelled the electromagnetic response of pigmy resonances with gain of insight into the symmetry energy.

\subsection{Nuclear Reactions in Stars}  

\subsubsection{\bf Reaction Rates and Radiative Capture}

In stars most nuclear reactions occur in the form of binary collisions. One of the exceptions is the triple-alpha reaction mentioned in the introduction.  
 For nuclei $j$\ and $k$\ in an astrophysical
plasma, the reaction rate in binary collisions is given by $r_{j,k}=\left<\sigma
v\right>n_{j}n_{k}$, where the reaction rate $\left<\sigma v\right>$\ is the average of the product of the cross section and the relative velocities, $\sigma v$,
over the temperature distribution.\ More
specifically,
\begin{equation}
r_{j,k}   =
\left<\sigma\mathrm{v}\right>_{j,k}{n_{j}n_{k}\over 1+\delta_{jk}},\label{astrophys4}
\end{equation}
where
\begin{equation}
 \left<\sigma\mathrm{v}\right>_{j,k}=\left({\frac{8}{{m_{jk}\pi}}}\right)^{1/2}%
(kT)^{-3/2}\int_{0}^{\infty}E\sigma(E)\exp\left(-{E\over kT}\right)dE.\label{astrophys5}%
\end{equation}
Here $m_{jk}$ denotes the reduced mass of the target-projectile system. The factor $1+\delta_{jk}$ in Eq. \eqref{astrophys4} 
accounts for the case of
identical particles, $i=k$. The rate should be
proportional to the number of pairs of interacting particles
in the volume.  If the particles are distinct, that is just
$ \propto n_j n_k $.
But if the particles are identical, the sum over distinct
pairs is
$ \propto n_j n_k/2 $.

In Eq. \eqref{astrophys5} the velocity, or kinetic energy $E$, distribution of the nuclei in the plasma at temperature $T$ is assumed to be Maxwellian. It has been shown in many instances that even a small depart from the Maxwell-Boltzmann velocity distribution would modify dramatically the outcome of the stellar evolution. For example, recently it has been shown that alternative statistics would ultimately have to be extremely close to Boltzmann-Gibbs statistics to reproduce the observed relative abundance of elements evolved during the big bang nucleosynthesis \cite{CFH13}. 

The direct measurement of the reaction cross sections $\sigma$ for nuclear astrophysics is difficult because it either involves neutron-induced reactions such as in the r-process, or charged particle reactions which are strongly inhibited by the Coulomb repulsion, or Coulomb barrier. Many of such reactions also involve nuclei far from the stability or neutron-rich nuclei. These nuclei are not usable as targets in laboratory experiments. Thus, most experiments with nuclei far from the stability requires the use of radioactive beam facilities. They are also done at much higher energies than the typical nuclear relative motion energy in the stars. Therefore, many indirect methods have been devised to extract indirectly the cross section for nuclear reactions in stars.

One of such indirect methods is known as the {\it Coulomb dissociation method}. This method is useful when a projectile is loosely-bound and can be dissociated ($a \rightarrow b+c$) in the Coulomb field of a target with a large charge $Z$.  The Coulomb breakup cross section for $a+A\longrightarrow b+c+A$ is given by \cite{BBR86,Ber88}, 
\begin{equation}
{d\sigma_{C}\over dE d\Omega}=\sum_{\Pi L}{dn^{\Pi L}\over dEd\Omega} \
\sigma_{\gamma+a\ \rightarrow\ b+c}^{\Pi L}(E),\label{CDmeth}
\end{equation}
where $\Pi = E$(electric) or $M$(magnetic),  $L=1,2,3, \cdots$ is the multipolarity, and  $\sigma_{\gamma+a\ \rightarrow\ b+c}^{\pi\lambda}(E)$ is the  the photonuclear cross section for  photon energy $E$.  $dN^{\Pi L}/dEd\Omega$ are the {\it equivalent photon numbers}, depending on $E$,  and the scattering angle $\Omega=(\theta,\phi)$ for $a+A\longrightarrow b+c+A$.
The equivalent photon numbers are obtained theoretically \cite{Ber88} for each
multipolarity ${\Pi L}$. Time reversal allows one to relate the radiative
capture cross section $b+c\longrightarrow a+\gamma$ to the photo-breakup cross section $\sigma_{\gamma+a\ \rightarrow\ b+c}%
^{\pi\lambda}(\omega)$. This method was proposed in Ref. \cite{BBR86} and has
been applied to several  reactions of interest for astrophysics.
For example,  the
reaction $^{7}$Be$($p$,\gamma)^{8}$B relevant for the production of high-energy solar neutrinos has been studied with this method \cite{Tohru,EBS05,Izs13}. 

For many systems studied via the Coulomb dissociation method, the contribution of the nuclear breakup is relevant and cannot be ignored. Such contributions has been examined by several
authors (see, e.g. \cite{BG98}).  In the case of  loosely-bound systems such as $^8$B it has
been shown that multiple-step, or higher-order effects, are
important \cite{BC92,BBK92,BB93} due to  continuum-continuum transitions. The fragments can make excursions in the continuum and iterate with the Coulomb field of the target  many times before they become asymptotically free. Multistep processes in the continuum led to the development of new techniques, such as the {\it continuum-discretized coupled-channels} (CDCC) method, e.g., as applied to a loosely-bound nucleus in Ref. \cite{BC92}. 
The effects of high multipolarities (such as E2 contributions) have also been studied \cite{Ber94,GB95,EB96} for the reaction $^{7}$Be$($p$,\gamma)^{8}$B). It has also been shown that the
influence of giant resonance states can be neglected \cite{Ber02}.  The challenge is to develop reliable nuclear models to obtain the scattering states including resonances. 

\subsubsection{\bf Single-Particle, Ab-Initio Methods} 

Taking  radiative capture reactions as an example, the cross sections for direct capture
are given by \cite{Ber94}
\begin{equation}
\sigma_{EL,J_{b}}^{\rm d.c.}   \propto  \left\vert \left\langle l_cj_c\left\Vert
\mathcal{O}_{\Pi L}\right\Vert l_{b}j_{b} \right\rangle
\right\vert^{2},
\label{respf}%
\end{equation}
where a proportionality factor involving phase-space quantities is omitted, $\mathcal{O}_{\Pi L}$ is the electromagnetic operator, and $\left\langle
l_cj_c\left\Vert \mathcal{O}_{\Pi L}\right\Vert l_{b}j_{b}
\right\rangle$ is a multipole matrix element involving bound ($b$) and continuum ($c$) wavefunctons. For 
electric multipole transitions  and in the long-wavelength approximation, $ \mathcal{O}_{\Pi L}= r^LY_{LM}$,
\begin{equation}
\left\langle l_cj_c\left\Vert
\mathcal{O}_{EL}\right\Vert l_{b} j_{b}\right\rangle
\propto   \int_{0}^{\infty}dr \
r^{L}u_{b}(r)u_{c}(r)
,\label{lol0}%
\end{equation}
where $u_i$ are radial single-particle wavefunctions. If the wavefunctions are obtained in a many-body model, including anti-symmetrization and correlations, the equations can get more involved. If a relation with the single-particle model can be made, then
the total direct capture cross section is obtained by adding all
multipolarities and final spins of the bound state ($E\equiv E_{nx}$),
\begin{equation}
\sigma^{{\rm d.c.}} (E)=\sum_{L,J_{b}} (SF)_{J_{b}}\ \sigma^{{\rm d.c.}%
}_{EL,J_{b}}(E) \ , \label{SFS}%
\end{equation}
where $(SF)_{J_{b}}$ are spectroscopic factors. The spectroscopic factors are adjusted to a shell model calculation for the single-particle occupancy amplitudes. This kind of calculations have been routinely used in the literature for the  $^{7}$Be$($p$,\gamma)^{8}$B) and other radiative capture reactions (see, e.g., Refs \cite{Rob73,WK81,EB95,Ber94,BB00,DT03,Jun10}). 

Radiative capture reactions using EFT techniques, such as those mentioned in an earlier section have been reported in e.g. Refs.  \cite{RH11,ZNP14a,ZNP14b}. The advantage of this approach is the predictive power of EFT which allows for increasing the precision of calculations as more terms in the effective theory expansion is included (the dots). One of the main problems of this method is the inclusion of many-body intrinsic features such as anti-symmetrization and medium modification of the interactions.   

The
{\it Resonating Group Method} (RGM) or the
{\it Generator Coordinate Method} (GCM) are able to include anti-symmetrization of bound and continuum states on equal foot. These are based on a set of coupled integro-differential equations of the form \cite{TLT78}
\begin{equation}
\sum_{\alpha'} \int d^3 r'
\left[
H^{AB}_{\alpha\alpha'}({\bf r,r'})-EN^{AB}_{\alpha\alpha'}({\bf r,r'})
\right]
g_{\alpha'}({\bf r'})=0,\label{RGM}
\end{equation}
where $H(N)({\bf r,r'})=\langle \Psi_A(\alpha,{\bf r})|H(1)|
\Psi_B(\alpha',{\bf r'}) \rangle$. Here, $H$ is the Hamiltonian for the
system of two nuclei (A and B) with the energy $E$, $\Psi_{A,B}$ is the wavefunction
of nucleus A (and B), and $g_{\alpha}({\bf r})$ is a function to be found by numerical
solution of Eq. \eqref{RGM}. This function describes the relative motion of A and B in channel
$\alpha$ with full antisymmetrization between nucleons of A and B. Such calculations have been carried out successfully 
for many reactions of astrophysical interest (see, e.g., Ref. \cite{DB94}). 

To the present date, the RGM is  the most practical tool to describe nuclear scattering and fusion of light nuclei at low energies with the inclusion of anti-symmetrization. It was introduced in Ref. \cite{HW53} and by solving it one can calculate cross sections for reactions involving cluster-like nuclei such as $^{12}$C($\alpha$,$\gamma$)$^{16}$O \cite{LK85}.  This reaction is relevant for cosmology.  It is thought that if the cross section for this reaction would be twice the adopted value,  a 25 solar masses star will not produce $^{20}$Ne since carbon burning would cease. An oxygen rich star is more likely to collapse into a black hole while carbon rich progenitor stars is more likely to leave behind a neutron star \cite{WW93}.  Can you imagine how important this nuclear reaction is for Cosmology?  This reaction determines the density of black holes and neutron stars scattered through the Universe.

Other techniques, beyond the single-particle and RGM models have been pursued. The ultimate goal is to calculate the cross sections from an {\it ab-initio}-type of calculations, i.e. a calculation that starts from the bare nucleon-nucleon interaction and evolves to solve the many-body nuclear problem, including the continuum. First attempts have been done in Refs. \cite{Nol01,NBC06} with ab-initio bound state wavefunctions calculated with the Quantum Monte-Carlo Shell Model or the No-Core Shell Model. In both cases the continuum wavefunction was derived by assuming a single-particle cluster model for the scattering states. Later, the RGM has been combined with  NSCM to calculate radiative capture reactions \cite{QN08,NRQ11}.  Another example is a calculation of the transfer reaction $^3$H(d,n)$^4$He
cross section to explain the experimental data obtained at the National Ignition Facility \cite{NQ12}.

\subsubsection{\bf Transfer, Trojan Horses, ANCs, and Surrogates} 

Although some direct measurements are possible, calculating transfer reactions  such as $^6$Li(d,$\alpha$)$^4$He  is relatively difficult. In general, the measurements also suffer from the same kind of difficulties we mentioned earlier. Therefore, some indirect methods have been developed. Among these we cite the 
{\it Trojan Horse Method} (THM). Because of off-shell effects a nucleus $x$ can be carried inside a trojan nucleus and brought to react at low energies with a heavy target. The heavier and faster trojan nucleus allows $x$ to overcome the Coulomb barrier \cite{Bau86,Spi11}. The basic idea is to extract the cross section in the low-energy region of a two-body reaction with significant astrophysical impact, 
$
a+x \longrightarrow c+C,
$
from a suitable three-body reaction,
$
a+b \longrightarrow s+c+C,
$ at high energies. An example is the study of the S-factor for the $^2$H(d,p)$^3$H reaction by using $^3$He breakup in the $^2$H($^3$He,pt)H reaction. 
The analysis of THM data is done by applying the well-known theoretical formalism of the quasi-free (QF) process.  One of the clusters in the incoming projectile is the participant, $x$ (the deuteron in the
given example), while the spectator $s$, ($^4$He or p) are only weakly involved in the process. These assumptions allows one to relate the cross section to the momentum distribution of  the inter-cluster ($x-s$) motion inside $b$

Trojan Horse (TH) reactions allow for the use of higher bombarding energies, $E_a$,  high enough
to overcome the Coulomb barrier of $a+x$ in the entrance channel of
the reaction. Then the effect of the Coulomb barrier and electron
screening effects become small.
The triple differential cross section
in the center of mass of the TH reaction can be written as
\begin{equation}
{d^3\sigma \over dE_cd\Omega_cd\Omega_C} = K_F \left|\Phi(p_{sx})\right|^2 \sum_{l_l} \left| L_{l_l}\right|^2 \left( {d\sigma_{l_l} \over d\Omega_{c.m.}}\right)^{HOES},
\label{HOES}
\end{equation}
where $l_l$ is the orbital angular momentum of particles $s$ and $x$ and $L_{l_l}$ is
a function of relative momentum and kinetic energy in the entrance channel  \cite{Mukunp}. In this equation,
$(d\sigma/d\Omega)_{cm})^{HOES}$ is the half-off-energy-shell (HOES) differential cross section for the two-body reaction at
the center-of-mass energy given  by
$
E_{cm} = E_{c-C} - Q_{2b}, 
$
where $Q_{2b}$ is the two-body $Q$-value of the binary process and $E_{c-C}$ is the relative energy between the outgoing
particles, $K_F$ is a kinematical factor, and $\Phi(p_{sx})$ is the Fourier transform of the radial wave function $\chi(r)$ for the $x-s$ intercluster motion.
The success of the THM relies on the QF kinematics (equivalent to $\Phi(p_{sx})\sim 0$ \cite{Che96,Zad89,Spi01,Lat01,Piz13}.

One has shown that the THM obtains is  the same cross section energy dependence as those obtained with direct methods \cite{Spi13}. The THM data are renormalized to those obtained with direct methods at high energies and the data at low energies are obtained as a bonus. The THM has been applied to a large number of reactions of interest for the production of light elements in numerous stellar environments to reactions of interest for the BBN \cite{Piz14}.  One of the many advantages of using transfer reaction techniques over direct measurements is to avoid the treatment of the {\it electron screening
problem} \cite{Con07,Mu10,Xu94}. This is just because the reaction is peripheral, probing the tails of the bound-state wave functions. In fact, most astrophysical nuclear fusion processes reed only this information for an accurate theoretical calculation.

The {\it Asymptotic Normalization Coefficients}
(ANC) method, introduced by Akram Mukhamedzhanov and collaborators \cite{Muk90,Tri06,Mu10}, assumes that the amplitude for
the radiative capture cross section $b+x\longrightarrow a+ \gamma$
is calculated according to 
\begin{equation}M=\left< I_{bx}^a({\bf r_{bx}})\Big| {\cal O}({\bf r_{bx}})\Big|
\psi_i^{(+)}({\bf r_{bx}})\right>,
\end{equation} where 
\begin{equation}I_{bx}^a=\left< \phi_a(\xi_b, \ \xi_x,\ {\bf 
r_{bx}}) \Big| \phi_x(\xi_x)\phi_b(\xi_b)\right>
\end{equation}
is the overlap function, i.e., an integration over the
internal coordinates $\xi_b$, and $\xi_x$, of $b$ and $x$. At low energies, $I_{bx}^a$ is
dominated by contributions of large $r_{bx}$. Thus,  the matrix element $M$ is dominated by the asymptotic
value of 
\begin{equation}I_{bx}^a\sim C_{bx}^a {1\over r_{bx}} W_{-\eta_a, 1/2}(2\kappa_{bx}
r_{bx}),
\end{equation} 
where  $W_{\alpha,n}$ is the
Whittaker function and $C_{bx}^a$ is the ANC. The ANC can be related to single-particle properties by writing it as
a product of the single particle (s.p.) spectroscopic factor and a normalization constant which depends on
the details of the s.p. wave function in the interior part of the
nucleus. If  the reaction occurs at very low energies, then the radiative capture cross section only probes the tails of the bound-state wave functions. Therefore, $C_{bx}^a$ is the only unknown part of the bound state wavefunction needed to calculate the direct capture cross section. The ANCs can be obtained by experimental analysis  of  elastic scattering between nuclei by extrapolation of the scattering phase shifts data to the bound state pole in the
energy plane. They can also be accessed in peripheral transfer reactions whose amplitudes
contain the same overlap function as the amplitude of the
corresponding astrophysical radiative capture cross section. 
 
As an example, consider the proton transfer reaction $
A(d,a)B$, where $d=a+p$, $B=A+p$. Using the asymptotic form of the
overlap integral, the cross section is given by
\begin{equation}
{d\sigma\over d\Omega} =
\sum_{J_Bj_d}\left[{(C_{Ap}^d)^2\over \beta^2_{Ap}}\right]\left[{(C_{ap}^d)^2\over \beta^2_{ap}}\right]
{\tilde \sigma}
\end{equation}
where $\tilde \sigma$ is a reduced cross section,  $\beta_{ap}$ ($\beta_{Ap}$) is the
asymptotic normalization of the shell-model bound-state proton wave
functions in nucleus $d (B)$, which are related to the corresponding
ANC's of the overlap function by means of $(C_{ap}^d)^2 =S^d_{ap}
\beta^2_{ap}$, where $S^d_{ap}$ is the spectroscopic factor. If the reaction $A(d,a)B$ is peripheral, the bound-state
wavefunctions used to calculate $\tilde \sigma$ are also approximated by their
asymptotic form, leading to $\tilde \sigma \propto \beta_{Ap}^2
\beta_{ap}^2$. Hence 
\begin{equation}
{d\sigma\over d\Omega} =
\sum_{j_i}(C_{Ap}^d)^2(C_{ap}^d)^2 R_{Bd}, \ \ \
{\rm where} \ \ \  R_{Bd}={{\tilde
\sigma}\over \beta^2_{Ap} \beta^2_{ap}}
\end{equation} 
is independent of $\beta^2_{Ap}$
and $\beta^2_{ap}$. Thus for surface-dominated reactions, the cross
section is obtained in terms of the product of
the square of the ANC's of the initial and the final nuclei $%
(C_{Ap}^d)^2(C_{ap}^d)^2$ rather than in terms of spectroscopic factors.

Most of the (n,$\gamma$) reactions of astrophysical interest will never be measured directly. Transfer reaction methods as the ones described above are form a kind of {\it surrogate reactions} in which the neutron is carried along by a nucleus to react with a target \cite{CB70,ED10}. An example of surrogate reactions are (d,p) reactions, which exploits the simplicity of the deuteron structure.  Evidently, the neutron in a nuclear environment carries angular momentum and other quantum numbers which are different than those of free neutrons. Unless the angular momentum plays a minimal role in the reaction, this brings  difficulties in extracting the desired neutron-induced reaction from the surrogate equivalent. Theoretically, this is the same as the implication that the Hauser-Feshbach formalism for (n,$\gamma$) reactions agrees with the Ewing-Weisskopf formalism \cite{CI10}. To elaborate on top of our discussion of the Hauser-Feshbach formalism leading to Eq. \eqref{HFd}, assume that
$\sigma_N(c)$ is the cross section for the formation of a {\it compound
nucleus} in the entrance channel $c$. Using the time-reversal theorem, the cross-section
$\sigma_{cc^{\prime}}$ can be related to the cross-section for the time-reversed
process $c^{\prime}\rightarrow c$, and one gets
\begin{eqnarray}
\sigma_{cc^{\prime}}\left(  E_{c^{\prime}}\right)
dE_{c^{\prime}}=\sigma _{CN}\left(  c\right) 
 \frac{\left( 2I_{c^{\prime}}+1\right)  \mu_{c^{\prime
}}E_{c^{\prime}}\;\sigma_{CN}\left(  c^{\prime}\right)  \omega(U_{c^{\prime}%
})dU_{c^{\prime}}}{\sum_{c}\int_{0}^{E_{c}^{\max}}\left(
2I_{c}+1\right)
\mu_{c}E_{c}\;\sigma_{CN}\left(  c\right)  \omega(U_{c})dU_{c}}%
\ ,\label{21.54b}%
\end{eqnarray}
where $I_{c}$ is the angular momentum and $\mu_{c}$ is the reduced
mass in channel $c$. Fragments emitted with energy in the range $E_{c^{\prime}}$ to $E_{c^{\prime}%
}+dE_{c^{\prime}}$ leave the residual nucleus with energy in the
range $U_{c^{\prime}}$ to $U_{c^{\prime}}+dU_{c^{\prime}}$ where
$U_{c^{\prime}}=E_{CN}-B_{c^{\prime}}-E_{c^{\prime}}$ and $E_{CN}$
and $B_{c^{\prime}}$ are the compound nucleus energy
and the binding energy of the fragment in the compound nucleus. Eq.
\eqref{21.54b} is the \textit{Weisskopf-Ewing} formula  \cite{WW40}. The {\it level density} is approximately given by
 $\omega(U)\propto\exp(U/T)$, so that the emitted fragments
in the Weisskopf-Ewing theory follow a Maxwell evaporation spectrum.

The Weisskopf-Ewing theory does not explicitly consider the conservation
of angular momentum and parity $J$ and parity $\pi$. This is described by the
\textit{Hauser-Feshbach} expression \cite{HF}
\begin{equation}
\sigma_{cc^{\prime}}=\frac{\pi}{k^{2}}\sum_{J}\frac{\left(
2J+1\right)
}{\left(  2i_{c}+1\right)  \left(  2I_{c}+1\right)  }\frac{\sum_{s,l}%
T_{l}(c)\sum_{s^{\prime},l^{\prime}}T_{l^{\prime}}(c^{\prime})}{\sum_{c}%
\sum_{s,l}T_{l}(c)}\ ,\label{21.73}%
\end{equation}
where $T_l$ are barrier transmission probabilities that depend on spin and parities. 
If the  (d,p) reaction is well reproduced by the Ewing-Weisskopf theory, this is a hint that the reaction is a good surrogate to obtain (n,$\gamma$) and (n,fission) cross sections \cite{Kes10}.  This means that the surrogate reaction does not depend on spin and parities and that the (d,p) and (n,$\gamma$) and (n,fission) reactions populate the same states.

Surrogate reactions involving charged particles, such as  (d,p) reactions  are complicated because they require a correct treatment of the interaction between its constituents to all orders. A popular tool to treat these systems is the {\it Alt-Grassberger-Sandhas} (AGS) \cite{AGS67} formalism.  They arise from the Faddeev formalism treating the system  $(a+b) + A$, and its rearrangements $a + (b+A)$ and $b+ (a+A)$. In each of these channels the three-particle scattering is described in terms of the transition operators $T_{\beta\alpha}$, where $\alpha$($\beta$) corresponds to a set of channel permutation combination. They obey the AGS equations \cite{AGS67} that are a system of coupled integral equations given by
\begin{equation}
T_{\beta\alpha}(z)=\tilde{\delta}_{\alpha\beta} G_0^{-1}(z)+\sum_{\nu=1}^3\tilde{\delta}_{\beta\nu}t_\nu(z) G_0(z)T_{\nu\alpha}(z) \ ,
\end{equation}
where  $\tilde{\delta}_{\beta\nu} =1-\delta_{\beta\nu}$ is the anti-delta-Kronecker symbol, $G_0 =(E+i_{\epsilon^+}-H_0)^{-1}$ is the resolvent of the three-free particle c.m. energy for the Hamiltonian $H_0$. The two-body transition operator $t_\nu$ for each interacting pair with inter-potential $v_\nu$ is obtained from the {\it Lippmann-Schwinger} equation
$t_\nu=v_\nu+v_\nu G_0 t_\nu$. The solution of these set of equations require some approximations (see, e.g., Ref. \cite{Muk14}) and one of the main problem is the accurate treatment of the Coulomb interaction. Some methods have been developed to handle the Coulomb interaction by means of screened potential, but it only works for relatively light nuclei because the solutions oscillate wildly at the asymptotic region for large charges \cite{DF09}.

During  {\it supernovae core collapse}, temperatures and densities are so high that electrons  are captured in nuclei reducing the electron fraction in the medium, and driving the nuclear composition to more neutron rich and heavier nuclei. As a consequence, nuclei with $N >40$ dominate matter composition for densities larger than a few $10^{10}$ g cm$^{-3}$.  Simulations indicate that electron capture stops at such densities and the capture becomes entirely due to free protons.  
In order to ensure that such simulations are correct, one needs to study  electron capture in a large number of nuclei. These studies cannot be carried out in a direct manner in the laboratory. The main object of interest are {\it Gamow-Teller matrix elements}, $B(GT)$, which cannot be extracted from $\beta$-decay experiments
The method of {\it charge exchange reactions} was developed to access the B(GT) matrix elements using (p,n), ($^3$He,t) and other charge exchange reactions at energies around 100 MeV/nucleon \cite{Sas12}. 
In such reactions one obtains the nuclear response by spin and isospin operators also involved in neutrino-scattering reactions occurring in stellar environments, useful to understand the neutrino driven explosion mechanism  \cite{MLD00}. 
This approach
relies on the similarity in spin-isospin space of charge-exchange
reactions, electron capture, neutrino scattering, and $\beta$ decay mechanisms. For example, 
the cross section $\sigma(p,n)$ at small momentum
transfer $q$ is given by \cite{Taddeucci1987}, 
\begin{equation}
{d\sigma\over dq}(q=0)=KN_D|J_{\sigma\tau}|^2 B(\alpha) , \label{tadeucci}
\end{equation}
where $K$ is a kinematical factor, $N_D$ is a distortion factor
(accounting for initial and final state interactions),
$J_{\sigma\tau}$ is the Fourier transform of the effective
nucleon-nucleon interaction, and $B(\alpha=F,GT)$ is the reduced
transition probability for non-spin-flip, $B(F)= (2J_i+1)^{-1}|
\langle f ||\sum_k  \tau_k^{(\pm)} || i \rangle |^2$, and spin-flip,
$B(GT)= (2J_i+1)^{-1}| \langle f ||\sum_k \sigma_k \tau_k^{(\pm)} ||
i \rangle |^2$, transitions. 

The 
charge-exchange matrix element is given by \cite{Ber93} 
\begin{equation}
{\mathcal M}_{exch}({\bf q})=\left<\Psi_a^{(f)} ({\bf r}_a)\Psi^{(f)}_b({\bf
    r}_b) \left| e^{-i{\bf q}\cdot {\bf r}_a } 
v_{exch}({\bf q})e^{i{\bf q}\cdot {\bf r}_b} \right| 
\Psi_a^{(i)} ({\bf r}_a)\Psi^{(i)}_b({\bf r}_b) \right>, \label{Mexch}
\end{equation}
where ${\bf q}$ is the momentum transfer, $\Psi_{a,b}^{(i,f)}$ are the intrinsic
wavefunctions of nuclei $a$ and $b$ for the initial and final states, ${\bf
  r}_{a,b}$ are the nucleon coordinates within $a$ and $b$, and $v_{exch}$ is
the part of the nucleon-nucleon interaction containing spin and isospin operators. 
For forward scattering and  low-momentum transfers,
${\bf q}\sim 0$, and the matrix element \eqref{Mexch} becomes
\begin{equation}
{\mathcal M}_{exch}({\bf q} \sim 0) \sim
v_{exch}^{(0)} ({\bf q}\sim 0) \, {\mathcal M}_a(F,GT)\,
{\mathcal M}_b(F,GT)
\, ,
\label{q1}
\end{equation}
where $v_{exch}^{(0)} $ idoes not depend on spin-isospin, and ${\mathcal M}_{exch}(F,GT)=
\left<\Psi_{a,b}^{(f)}\vert\vert (1 \ {\rm or} \ \sigma ) \tau \vert\vert
  \Psi_{a,b}^{(i)}\right>$ are Fermi or Gamow-Teller (GT) matrix 
elements for the nuclear transition.
A similar result can be obtained using eikonal
scattering waves instead of plane waves. This justifies the use of Taddeucci's formula, as long as
one remains within the validity of the
low-momentum transfer assumption in high-energy collisions, as discussed thoroughly in Ref. \cite{Ber93}.  
The proportionality between cross section and Gamow-Teller strength was shown to apply to a wide range of  mass numbers \cite{Zegers2007}. 

\subsubsection{\bf Hadrons, Quark Soups, and Strong E/M Fields}

It is predicted that during the Big Bang, a phase transition occurred from a soup of quarks and gluons to a hadronic phase. This soup was present in an environment of high temperature and densities. A phase transition from hadrons to quarks and gluons, although in a cold environment, might also occur in the core of neutron stars. The only possible way to study phase transitions to sub-nucleon degrees of freedom are by means of  ultra-relativistic heavy ion collisions (URHIC), routinely carried out at the RHIC accelerator in Brookhaven and at the LHC
at CERN to induce very high temperatures, densities, and pressures in central collisions.
Under such conditions, it is expected that the nucleons are excited into baryonic
resonance states, along with particle production creating a hadronic resonance matter and a possible transition
to a free soup of quarks and gluons during a short time \cite{wong,HW04}.  QCD induced   phenomena, such as {\it quark confinement}  and {\it spontaneous chiral symmetry breaking}  are  caused by infrared physics based on {\it asymptotic freedom}, but  it is not fully understood how they are related with each other. LQCD simulations predict that  these two phenomena are apparently turned off at about the same time at a given temperature at the onset of the QGP \cite{Kog83,Baz13,Fodo09,Miya95,Wol95}.

By studying URHICs one extracts of the energy density
$\epsilon$, pressure $P$ , and entropy density $s$ of the
soup as a function of the temperature $T$ and the
baryochemical potential $\mu_{B}$. During a phase transition to a {\it quark-gluon plasma} (QGP)  a rapid rise in
the effective number of degrees of freedom, expressed by
$\epsilon/T^{4}$, or $s/T^{3}$, is observed. The variables $T$, $s$, and $\epsilon$, are correlated
with the average transverse momentum $\left\langle
p_{T}\right\rangle $, the hadron rapidity density $dN/dy$, and the
transverse energy density $dE_{T}/dy$. The transverse
energy produced in the interaction is given by $
E_{T}=\sum_{i}E_{i}\sin\theta_{i}\ , $ where $E_{i}$ and
$\theta_{i}$\ are the kinetic energies of the fragments and their
emission angles. Photons and leptons produced in URHIC provide
information on the various stages of the interaction without
modification by final state interactions. The widths and masses of the mesons $\rho$,
$\omega$, and $\phi$ appearing in the lepton pair invariant mass
spectrum are expected to be sensitive to medium effects. J/$\Psi$ particles are abundantly produced in a quark-gluon plasma and predicted to be suppressed due to  the \textit{Debye screening} of the $c\overline{c}$ pairs. Loosely-bound
states of the $c\overline{c}$ system, such as
$\Psi^{\prime}$ and $\chi_{c}$ , are more easily dissociated and
are even more  suppressed than the J/$\Psi$.
In a QGP the strange quark content is rapidly saturated
by $s\overline{s}$ production in gluon-gluon interactions. This leads to
an enhancement in the production multi-strange baryons and strange antibaryons when a QGP is formed.

The energy loss of a quark is related to the color-dielectric polarizability of the medium, in
analogy with the energy loss by electromagnetic probes. Due to Bremsstrahlung,  radiation is a very efficient energy loss mechanism for relativistic particles. But it is strongly suppressed in a dense medium by the
\textit{Landau-Pomeranchuk effect} \cite{wong}. By analogy, the {\it stopping power} of particles within a QGP is predicted to be higher than that in the hadronic matter, leading to an attenuation the emission of {\it gluon jet} pairs from parton-parton collisions in the direction opposite to the trigger jet, known as jet-quenching and visible in measurements of high-p$_T$ spectra. A quark or
gluon jet propagating through a dense medium will not only
loose energy but will also be deflected, leading to
an azimuthal asymmetry of the jets \cite{Agg00,Bat06,Dam06,Ada06,Bue06}.

In order to understand the QGP phase transitions, the boundary conditions in the collisions have to be understood properly. For example, the {\it parton distribution functions} (PDF) which are basically medium modified gluon distributions in nuclei have to be studied carefully.  The strong electromagnetic field generated by one of the colliding ions has been proposed in Ref. \cite{GB02} to be used as a tool to study PDFs at the LHC/CERN. The physics of ultra-peripheral collisions is reviewed in Refs. \cite{BB88,BKN05,Bal08}. The sensitivity of the different modern PDFs to the medium corrections is clearly visible in the rapidity spectrum of J/$\Psi$ and $\Upsilon$ particles produced via this process \cite{AB11,AN13,Reby12}. The power of experiments to discern among the several PDFs \cite{AB11} those that best reproduce the experimental data has been evidenced in experiments at the LHC/CERN \cite{Abe13,AB13,expups}. This allows for a better discrimination of the pre-initial PDF conditions for interest in the formation of the QGP in URHICs.

The experimental discovery of a QGP was announced in different occasions in the last few decades \cite{SZ86,CERN00,Gyu04,BNL10,Nat12}.  The numerous press releases with  announcements of the QGP discovery reflects the extreme difficulty to produce and confirm the existence of such an ephemerally state  of nuclear matter, lasting an extremely short time with temperatures as high as $10^{12}$ K. Of present interest is also  the {\it glasma state} of matter occurring when each of the two nuclei in URHIC form a color glass condensate, related to the PDFs, preceding the QGP.  In such a state the dressed partons are condensed into a glassy state, prior to the transition between the confined state within hadrons and QGP state \cite{IV03,LM06,GV06}.

After the experimental discovery of the Higgs boson \cite{CMS12,ATLAS12}, the CERN experimentalists are searching for new phenomena beyond the standard model of particle physics, among others there are dedicated searches for supersymmetric particles \cite{GS71,Vol72,Ram71,WZ74,Mar97} and extra-dimensions \cite{Kal21,Ark98,App01,Rizz01,McM02}. But there is also a very strong community at CERN studying nuclear physics, not only with regard to the formation and decay of the QGP, but also for other purposes. At present, exciting experiments at much lower energies are being carried out in nuclear physics  at the CERN/ISOLDE radioactive ion beam facility \cite{isolde} and many other sort of breakthrough experiments in nuclear physics make the headlines now and then. A few examples are the early production of {\it anti-hydrogen}, the first time and anti-atom was produced and detected \cite{Bau96} (see also \cite{BeBa88,Blan97,BeBa98}, or the recent precision measurements of the mass difference between light nuclei and {\it anti-nuclei} \cite{AL15}.

\section{Dreams Come True}

One of the dreams of human technology is to reproduce on Earth the same fusion processes occurring in the Sun, namely, the immense energy release through fusion of light nuclei. We have developed Tokamaks, marvelous machines that confine the hot plasma necessary to trigger nuclear fusion. However, we have never been able to develop a machine that produces more fusion energy than it is employed to run it. It is a pity, as fusion energy is free from many of the problems related to nuclear reactors, most notably the immense nuclear radiation problem.  Large multinational projects seem to be the only way to couple with such incredibly expensive tasks. We are right now building a fantastic Tokamak machine in France, with funds from several countries flowing in. The project is very costly, possibly reaching 20 billion Euros or more by the end of its completion. As of today, the machine is  predicted to obtain the first plasma by 2030 \cite{ITER}. Some of us will not be here to witness it.

Unfortunately, ITER (pronounced ``eater"), an acronym for International Thermonuclear Experimental Reactor, is running into trouble, as countries funding the project do not deliver their sub-projects on time, delaying the programmed deadlines. A similar case, at much smaller funding scale, is going on with the FAIR facility in Germany \cite{fairf}. It seems that the best approach for expensive multinational projects is to allow a single country to run most (if not all) of the project steps, much like the CERN model has proven to function with success. From the science point of view, ITER does not plan much of nuclear science research. Basically the project is related to material science and the studies of what radiation can do to the material confining the plasma. A new ITER possibly built many decades after (we will all be dead), might then reach the so much expected fusion energy gain.  This is sort of disappointing. But it might also be a fallacy of the way we think about  such projects. Who would predict 3 decades ago that studies of anti-hydrogen might be of interest for the CERN laboratory?, or that tests of ab-initio nuclear structure calculations might be of interest for the National Ignition Facility (NIF)? In the end, nuclear physicists will certainly find their niche in the ITER project, just because nuclear physics research is resilient and necessary.

Many other dreams are out there, some of them perfectly doable within present funding capabilities, or already being accomplished. Among a few, we cite heroic efforts by Toshimi Suda leading  SCRIT \cite{Suda}, a novel experimental technique to trap short-lived nuclei almost at rest, a  completed electron scattering facility inside RIKEN/RIBF employing a recycled electron storage ring. First experiments are underway. Also worth mentioning are the efforts of Raju Venugopalan, Abhay Deshpande, and collaborators to promote the idea of an Electron-Ion Collider (EIC) in the U.S. using extensions of present facilities.  Such a facility would probe the structure of nuclei down to distance scales as small as $10^{-3}$ fm  allowing the study of momentum, spatial, spin and orbital distributions of gluons and sea quarks in light and heavy nuclei \cite{Raju15} (see also \cite{Abel12}). 

Many similar ideas are flourishing in nuclear physics and many of them are found to be extremely useful for the progress of science.

\section{Conclusions, Excuses and Complaints}
In this review we have attempted to convey a realistic picture of the current status of nuclear physics research. The choice of the topics covered in the review reflects our own interest and contributions to the field. Clearly several topics were left out. The message which we hope to transmit is that the field is quite alive with future planning and developments are underway and will guarantee a thriving field of physics research in the next decades. Nuclear physics is a synergetic field in the sense that ideas, theories, and concepts find a natural application in other fields, such as cold atomic gases (Feshbach Resonances), atomic and molecular collisions (CDCC), and universal conductance fluctuations in mesoscopic systems such as open quantum dots and grapheme flakes (Quantum Chaotic Scattering and RMT). 

In the review we have also discussed the recent development in the reaction theory used to treat the collisions of weakly bound and halo nuclei. The fusion and incomplete fusion of these systems are discussed and the theory employed robustly analyzed. We have also discussed the relevance of nuclear physics to astrophysics and presented a rather detailed account of the type of reactions of importance to nucleosynthesis. The underlying QCD and its constraints on nuclear phenomena were also discussed. The quark-gluon plasma predicted to exist, albeit for a very short time, in the high LHC energy collisions of heavy ions is also briefly discussed.

In summary, in this review we have tried to give a flavor of what Nuclear Physics means for science, and what are its most actual academic issues.  There are more facets to Nuclear Physics than we can cover in such a short review and we had to leave out the discussion of a  large number of other interesting research topics.  

\bigskip

\begin{acknowledgments}
CAB is supported by the U.S. NSF Grant No. 1415656, and U.S. DOE grant No. DE-FG02- 08ER41533 and by the Fulbright U.S. Scholar Program.
 MSH is partially supported by the Brazilian funding agencies, FAPESP, CNPq, CAPES/ITA-PVS Fellowship Program, The INCT-IQ/MCT, and CEPID-CEPOF.
\end{acknowledgments}


\begin{thebibliography}{99}
\bibitem{Ma49} M. G. Mayer, Phys. Rev. 75, 1969 (1949); Phys. Rev. 78,  16 (1950).
\bibitem{Ha49} O. Haxel, J. H. D. Jensen, and H. E. Suess, Phys. Rev. 75, 1766 (1949).
\bibitem{BM69} A. Bohr and B. Mottelson, ``Nuclear Structure", Vols. I and II Benjamin, New York (1969).
\bibitem{BCS} J. Bardeen, L.N. Cooper, J.R.  Schrieffer,  Phys. Rev. 106, 162 and 108, 1175 (1957).
\bibitem{Weisskopf} J. M. Blatt and V. F. Weisskopf, Theoretical Nuclear Physics, J. Wiley and Sons, New York (1952).
\bibitem{Feshbach1993} H. Feshbach, ``Theoretical Nuclear Physics: Nuclear Reactions", Wiley-VCH (1993).
\bibitem{Tal93} I. Talmi, ``Simple Models of Complex Nuclei: The Shell Model and the Interacting Boson Model", Harwood Academic Publishers (1993).
\bibitem {wong} C.Y. Wong,  ``Introduction to high energy heavy ion collisions", World Scientific, Singapore  (1994).
\bibitem{Sat83} G.R. Satchler, ``Direct Nuclear Reactions", Clarendon Press (1983).
\bibitem{BD04} C.A. Bertulani and P. Danielewicz, ``Introduction to Nuclear Reactions", IOP (Taylor and Francis), London (2004).
\bibitem{RS04} P. Ring and P. Schuck, ``The nuclear Many-Body Problem", Springer Science \& Business Media (2004).  
\bibitem{Bertu07} C.A. Bertulani, ``Nuclear Physics in a Nutshell", Princeton University Press, Princeton, NJ (2007).
\bibitem{CH13}L. F. Canto and M. S. Hussein, ``Scattering Theory of Molecules, Atoms and Nuclei", World Scientific (2013).
\bibitem{BK47} G. Baldwin, and G. Klaiber,  Phys. Rev. 71, 3 (1947).
\bibitem{SJ50} H. Steinwedel,  H.D.Z. Jensen,  Naturforsch. 5a, 413 (1950).
\bibitem{GT48} M. Goldhaber and E.  Teller, Phys. Rev. 74, 1046 (1948).
\bibitem{Berm75} B.L. Berman, and S.C.  Fultz,  Rev. Mod. Phys. 47, 713 (1975).
\bibitem{Bert76} F.E.  Bertrand,  Ann. Rev. Nucl. Sci. 26, 457 (1976).
\bibitem{Spet81} J. Speth, and A. Van der Woude,  Rep. Prog. Phys. 44, 719 (1981).
\bibitem{AI75} A. Arima and F. Iachello, Phys. Rev. Lett. 35, 1069 (1975).
\bibitem{AI81}  Arima and F. Iachello, ``The Interacting Boson Model", Ann. Rev. Nucl. Part. Sci. 31, 75 (1981).
\bibitem{Bohr} N. Bohr, Nature 137, 344 (1936).
\bibitem{Fes58a} H. Feshbach,  Ann. Rev. Nucl. Sci. 8, 49 (1958).
\bibitem{KRY1963}A. K. Kerman, L. S. Rodberg  and J. E. Young, Phys. Rev. Lett. 11, 422 (1963).
\bibitem{BF1963} B. Bloch and H. Feshbach, Ann. Phys. (NY) 23, 47 (1963).
\bibitem{FKL1967} H. Feshbach, A. K. Kerman and R. H. Lemmer, Ann. Phys. (NY), 41, 230 (1967).
\bibitem{Work1996} Int. Workshop on Physics of Unstable Nuclear Beams, Serra Negra, Brazil, 1996, ed. C.A. Bertulani, L.F. Canto and M.S. Hussein, World Scientific, Singapore (1996).
\bibitem{Work1998} Int. Workshop on Collective Excitations in Fermi and Bose Systems, Serra Negra, Brazil, 1998, ed. C.A. Bertulani, L.F. Canto and M.S. Hussein, World Scientific, Singapore (1998).
\bibitem{Bet39} H.A. Bethe,  ``Energy Production in Stars", Phys. Rev. 55, 434 (1939).
\bibitem{Ad11}`E.G. Adelberger et. al., `Solar fusion cross sections II: the pp chain and CNO cycles",  Rev. Mod. Phys. 83, 195 (2011).
\bibitem{Gam28} G. Gamow, Z. Physik 51, 204 (1928). 
\bibitem{GC28}R. W. Gurney and E. U. Condon, Nature 122, 439 (1928).
\bibitem{Dav14} Alfonso F. Davila and Christopher P.  McKay,  Astrobiology 14, 534 (2014). 
\bibitem{Hoy54} F. Hoyle, Astrophys. J. Suppl. 1, 121 (1954).
\bibitem{Fow84} William A. Fowler, ``The quest for the origin of the elements", Science 226, 922 (1984).
\bibitem{SW87} S. Weinberg, Phys. Rev. Lett. 59, 2607 (1987).
\bibitem{Hjo11} M. Hjorth-Jensen,   Physics 4, 38 (2011). doi:10.1103/Physics.4.38.
\bibitem{Mei15} U.-G. Meissner, ``Anthropic considerations in nuclear physics", Science Bulletin, 1, 60 (2015). 
\bibitem{Fyn05} H.O.U. Fynbo et al., Nature 433, 136 (2005).
\bibitem{Bet49} H.A. Bethe, Phys. Rev. 76, 38 (1949).
\bibitem{Hi35} Hideki Yukawa, Proceedings of the Physico-Mathematical Society of Japan 17, 48 (1935).
\bibitem{Gel95} M. Gell-Mann, ``The Quark and the Jaguar", Owl Books (1995). 
\bibitem{PS95} M. E. Peskin and D. V. Schroeder, ``An introduction to quantum field theory". Addison-Wesley (1995).
\bibitem{PDPG} Particle Data Group Collaboration, J. Beringer et al., ``Review of particle physics," Phys. Rev. D 86, 010001 (2012).
\bibitem{BalS92}  G. Bali and K. Schilling,  Phys.Rev. D46, 2636 (1992).
\bibitem{Lepa05} G. Peter Lepage, arXiv:hep-lat/0506036 (2005).
\bibitem{FH12} Zoltan Fodor and Christian Hoelbling, Rev. Mod. Phys. 84, 449  (2012).
\bibitem{usqcd} http://www.usqcd.org
\bibitem{Berk15} Evan Berkowitz, Thorsten Kurth, Amy Nicholson, Balint Joo, Enrico Rinaldi, Mark Strother, Pavlos M. Vranas, Andre Walker-Loud, arXiv:1508.00886 (2015).
\bibitem{Sav15} Martin J. Savage, arXiv:1508.00892 (2015).
\bibitem{Wei79} S. Weinberg, Physica A96, 327 (1979).
\bibitem{ORK94} C. Ordonez, L. Ray, U. van Kolck, Phys. Rev. Lett. 72, 1982 (1994).
\bibitem{OK92} C. Ordonez, U. van Kolck, Phys. Lett. B291, 459 (1992).
\bibitem{Kolc94} U. van Kolck, Phys. Rev. C49, 2932 (1994).
\bibitem{ORK95} C. Ordonez, L. Ray, U. van Kolck, Phys. Rev. C53, 2086 (1996).
\bibitem{Lepa97} Peter Lepage, ``How to renormalize the Schr\"odinger Equation", Lectures given at the VIII Jorge Andre Swieca Summer School (Brazil, 1997), Ed. by C.A. Bertulani, M. Bracco, B. Carlson and M. Nielsen, World Scientific, Singapore (1998).
\bibitem{Kolc98} U. van Kolck, Nucl. Phys. A631, 56c (1998).
\bibitem{Bed98} Paulo F. Bedaque, U. van Kolck, Phys. Lett. B428, 221  (1998).
\bibitem{BedU99} Paulo F. Bedaque, H.W. Hammer, U. van Kolck, Nucl. Phys. A676, 357 (2000).
\bibitem{Barn15} N. Barnea, L. Contessi, D. Gazit, F. Pederiva, U. van Kolck, Phys. Rev. Lett. 114, 052501 (2015). 
\bibitem{Rup00} G. Rupak, Nucl. Phys. A678, 405 (2000).
\bibitem{BHK02} C.A. Bertulani, H.-W. Hammer and U. van Kolck, Nucl. Phys. A 712, 37 (2002).
\bibitem{Rup11} G. Rupak and R. Higa, Phys. Rev. Lett. 106, 222501 (2011).
\bibitem{Rup13} G. Rupak and D. Lee, Phys. Rev. Lett. 111, 032502 (2013).
\bibitem{Lee09} D. Lee, Prog. Part. Nucl. Phys. 63, 117 (2009).
\bibitem{Epe11} E. Epelbaum, H. Krebs, D. Lee, U.G. Mei$\beta$ner Phys. Rev. Lett 106, 192501 (2011).
\bibitem{Epel13} E. Epelbaum, H. Krebs, T. A. L\"ahde, D. Lee, U.-G. Mei$\beta$ner, Eur. Phys. J. A 49:82 (2013).
\bibitem{Epel14} E. Epelbaum, H. Krebs, T.A. L\"ahde, D. Lee, U.-G. Mei$\beta$ner, G.  Rupak, Phys. Rev. Lett. 112, 102501 (2014).
\bibitem{Lan14} T.A. L\"ahde, E Epelbaum, H Krebs, D Lee, U.-G. Mei$\beta$ner, G Rupak, Phys. Lett. B 732, 110 (2014).
\bibitem{Du08} S. D\"urr,  et al.,  Science 322, 1224 (2008).
\bibitem{Nog00} A. Nogga, H. Kamada and W. Gl\"ockle, Phys. Rev. Lett. 85, 944 (2000).
\bibitem{Schw02} A. Schwenk, G.E. Brown and B. Friman, Nucl. Phys. A703,  745 (2002).
\bibitem{Bog02b} S.K. Bogner et al., Phys. Rev. C65, 051301(2002).
\bibitem{Bog03} S.K. Bogner et al., Phys. Lett. B576, 265 (2003).
\bibitem{BogPR03} S.K. Bogner, T.T.S. Kuo and A. Schwenk, Phys. Rep. 386, 1 (2003) .
\bibitem{Schw03} A. Schwenk, B. Friman and G.E. Brown, Nucl. Phys. A713, 191 (2003).
\bibitem{Schw04} A. Schwenk and B. Friman, Phys. Rev. Lett. 92, 082501 (2004).
\bibitem{Stec10} I. Stetcu, J. Rotureau, B.R. Barrett, U. van Kolck, J. Phys. G37,  064033 (2010). 
\bibitem{Roth11} R. Roth, J. Langhammer, A. Calci, S. Binder, P. Navr\'atil, Phys. Rev. Lett. 107, 072501 (2011).
\bibitem{PBM07} Peter Braun-Munzinger and Johanna Stachel, Nature 448, 302 (2007).
\bibitem{Fes58} Herman Feshbach, Ann. Phys. (N.Y.) 5, 357 (1958).
\bibitem{Bloc05} Immanuel Bloch,  Nature Physics 1, 23 (2005). doi:10.1038/nphys138.
\bibitem{Bose24} S.N. Bose,  Zeit. Phys. 26,  178. (1924).
\bibitem{Einst25} A. Einstein, ``Quantentheorie des einatomigen idealen Gases". Sitzungsberichte der Preussischen Akademie der Wissenschaften 1, 3 (1925).
\bibitem{Aur07} A. Bulgac,  Phys. Rev. A 76, 040502 (2007).
\bibitem{Forb12} Michael Neil Forbes, ``The unitary fermi gas: an overview", INT, University of Washington, unpublished (2012). Available at
http://www.int.washington.edu/PROGRAMS/12-2c/week3/forbes.pdf
\bibitem{Bul11} Aurel Bulgac,  Yuan-Lung Luo, Piotr Magierski, Kenneth J. Roche, and Yongle Yu,  Science 332, 1288 (2011).
\bibitem{Bul14} Aurel Bulgac,  Michael McNeil Forbes, Michelle M. Kelley, Kenneth J. Roche, and Gabriel Wlazowski,  Phys. Rev. Lett. 112, 025301 (2014).
\bibitem{NNDC} National Nuclear Database Center, Brookhaven National Laboratory, http://www.nndc.bnl.gov
\bibitem{BCH93} C.A. Bertulani, L.F. Canto and M.S. Hussein,  Phys. Rep. 226, 281 (1993).
\bibitem{Bertulani02} C. A. Bertulani, M. S. Hussein, and G. M\"unzenberg, ``Physics of Radioactive Beams", Nova Science Publishing (NY) (2001).
\bibitem{Lic03} R. Lichtenth\"aler, A. L\'epine-Szily, V. Guimar\~aes, G.F. Lima, and M.S. Hussein. Braz. J. Phys.  33,  294 (2003).
\bibitem{LLG14} A. L\'epine-Szily, R. Lichtenthaler, and V. Guimar\~aes,  Eur. Phys. J. A,  50,128 (2014).
\bibitem{Fus11} C.A. Bertulani, ``Summary Talk - Fusion 2011", St. Malo, France,  EPJ Web of Conferences 17, 15001 (2011).
\bibitem{Sob66}A. Sobiczewski, F. A. Gareev, and B. N. Kalinkin, Phys. Lett. 22, 500 (1966).
\bibitem{Mye66}W. D. Myers and W. J. Swiatecki, Nucl. Phys. 81, 1 (1966).
\bibitem{Hof00}S. Hofmann and G. M\"unzenberg, Rev. Mod. Phys. 72, 733 (2000).
\bibitem{Cwi05}S. Cwiok, P.-H. Heenen, and W. Nazarewicz, Nature  433, 705 (2005).
\bibitem{Sob07}A. Sobiczewski and K. Pomorski, Prog. Part. Nucl. Phys. 58, 292 (2007).
\bibitem{Oga07}Yu. Ts. Oganessian, J. Phys. G 34, R165 (2007).
\bibitem{Hof07}S. Hofmann et al., Eur. Phys. J. A 32, 251 (2007).
\bibitem{Oga11}Yu. Ts. Oganessian, Radiochim. Acta 99, 429 (2011).
\bibitem{Khu14} J. Khuyagbaatar et al., Phys. Rev. Lett. 112, 172501 (2014).
\bibitem{BV72} D. Vautherin and D. Brink, Phys. Rev. C 5, 626 (1972).
\bibitem{Schu15} N. Schunck, J.D. McDonnell, J. Sarich, S.M. Wild, and D. Higdon, J. Phys. G: Nucl. Part. Phys. 42, 034024 (2015)
\bibitem{Eri12} J. Erler, N. Birge, M. Kortelainen, W. Nazarewicz, E. Olsen, A.M. Perhac, and M. Stoitsov, Nature 486, 509 (2012). 
\bibitem{Ben03} Michael Bender, Paul-Henri Heenen, and Paul-Gerhard Reinhard, Rev. Mod. Phys. 75, 121 (2003).
\bibitem{Ber09} G. F. Bertsch, C. A. Bertulani, W. Nazarewicz, N. Schunck, and M. V. Stoitsov, Phys. Rev. C 79, 034306 (2009).
\bibitem{GC14} S. Goriely and R. Capote, Phys. Rev. C 89, 054318  (2014).
\bibitem{Du14} T. Duguet, Lec. Notes Phys. 879, 293 (2014).
\bibitem{Cha15} N. Chamel, J.M. Pearson, A.F. Fantina, C. Ducoin, S. Goriely, A. Pastore, Acta Phys. Pol. 46, 249 (2015).
\bibitem{Nik11} T. Niksi\'c, ,D. Vretenar, P.Ring, Prog. Part. Nucl. Phys. 66, 519 (2011).
\bibitem{Zha15} P.W. Zhao,  N. Itagaki, J.  Meng, J. Phys. Rev. Lett. 115, 022501 (2015).  
\bibitem{Lian15}  Haozhao Liang, Jie Meng, Shan-Gui Zhou,   Phys. Rep. 570, 1 (2015). 
\bibitem{Meng06a} J. Meng, J. Peng, S. Q. Zhang, and S.-G. Zhou, Phys. Rev. C 73, 037303 (2006).
\bibitem{Meng06}  J. Meng, H. Toki, S.-G. Zhou, S. Q. Zhang, W. H. Long, and L. S. Geng, Prog. Part. Nucl. Phys. 57, 470  (2006). 
\bibitem{Lian08} H. Z. Liang, N. Van Giai, and J. Meng, Phys. Rev. Lett. 101, 122502  (2008).  
\bibitem{Zhou03}  S.-G. Zhou, J. Meng, and P. Ring , Phys. Rev. Lett. 91, 262501 (2003). 
\bibitem{Meng96} J. Meng and P. Ring, Phys. Rev. Lett. 77,  3963 (1996).
\bibitem{Meng98} J. Meng and P. Ring, Phys. Rev. Lett. 80,  460 (1998). 
\bibitem{Zhao11}  P. W. Zhao, J. Peng, H. Z. Liang, P. Ring, and J. Meng, Phys. Rev. Lett. 107,  122501 (2011). 
\bibitem{Lal04} G. A. Lalazissis, Peter Ring, D. Vretenar, ``Extended Density Functionals in Nuclear Structure Physics", Springer (2004).
\bibitem{AB13} P. Avogadro, C. A. Bertulani, Phys. Rev. C 88, 044319 (2013).
\bibitem{Sim10} C. Simenel, Phys. Rev. Lett. 105, 192701 (2010).
\bibitem{Bul11b} I. Stetcu, A. Bulgac, P. Magierski, and K.J. Roche,  Phys. Rev. C 84, 051309(R) (2011),
\bibitem{Bul13} Aurel Bulgac, Annu. Rev. Nucl. Part. Sci. 63, 97 (2013).
\bibitem{Kaz13} Kazuyuki Sekizawa and Kazuhiro Yabana, Phys. Rev. C 88, 014614 (2013).
\bibitem{Stet15} I. Stetcu, C. Bertulani, A. Bulgac, P.Magierski, and K.J. Roche, Phys. Rev. Lett. 114, 012701 (2015).
\bibitem{Tani96} Isao Tanihata, J. Phys. G 22, 157 (1996).
\bibitem{SagH15} Hiroyuki Sagawa and Kouichi Hagino, Eur. Phys. J. A 51, 102 (2015).
\bibitem{Bau10} E. Bauer, A.P. Gale\~ao, M.S. Hussein et al., Nucl. Phys. A 834, 599c (2010).
\bibitem{Wrob04} Andrzej K. Wroblewski, Acta Physica Polonica 35, 1 (2004).
\bibitem{ELI} Extreme Light Infrastructure, http://www.extreme.light.infrastructure.eu.
\bibitem{IZEST} International Center on Zetta-Exawatt Science and Technology, http://www.izest.polytechnique.edu.
\bibitem{Ber13} C.A. Bertulani, ``Nuclear Reactions",   ``Encyclopedia of Nuclear Physics and its Applications", Ch. 2, Edited by Reinhard Stock, Wiley (2013).
\bibitem{FPW54} H. Feshbach, C. E. Porter, and V. F. Weisskopf. Phys. Rev. 96, 448 (1954).
\bibitem{Huss00} M.S. Hussein, M. Ueda, A.J. Sargeant, M.P.  Pato, Phys. Rev. C 61, 45801 (2000).
\bibitem{SPP1} L. C. Chamon, D. Pereira, M. S. Hussein, M. A. C\^andido Ribeiro, and D. Galetti, Phys. Rev. Lett. 79, 5218 (1997).
\bibitem{SPP2} L. C. Chamon, B. V. Carlson, L. R. Gasques, D. Pereira, C. De Conti, M. A. G. Alvarez, M. S. Hussein, M. A. C\^andido Ribeiro, E. S. Rossi, Jr., and C. P. Silva, Phys. Rev. C 66, 014610 (2002).
\bibitem{HBC85} M.S. Hussein, A.J. Baltz, and B.V. Carlson,  Phys. Rep. 113, 133 (1985).
\bibitem{Gome06} L.C. Chamon, P.R.S. Gomes, M.S. Hussein, Phys. Rev. C 73, 077610 (2006).
\bibitem{FH80} W.E. Frahn,  M.S. Hussein,  Nuclear Physics A 346, 237 (1980).
\bibitem{HF} W. Hauser and H. Feshbach, Phys. Rev. 87, 366 (1952).
\bibitem{WW40} V.F. Weisskopf and D.H. Ewing, Phys. Rev. 54, 472 (1940).
\bibitem{KKM79} M. Kawai, A. K. Kerman and K. W. McVoy, Ann. of Phys. 75, 156 (1973).
\bibitem{FKK} H. Feshbach, A. K. Kerman, and S. E. Koonin, Ann. Phys. (NY) 125, 429(1980).
\bibitem{CEH} B. V. Carlson, J. E. Escher, and M. S. Hussein, J. Phys. G 41, 094003 (2014).
\bibitem{Feshbach58} H. Feshbach, Ann. Phys. (NY),  5, 357 (1958).
\bibitem{Feshbach62} H. Feshbach, Ann. Phys. (NY), 19, 287 (1962).
\bibitem{Eric60} T. Ericson, Phys. Rev. Lett., 5, 430 (1960).
\bibitem{Eric63} T. Ericson, Ann. Phys. (NY), 23, 390 (1963).
\bibitem{Breit95} A.M.S. Breitschaft, L.F.  Canto, M.S. Hussein, E.J. Moniz, J. Christley, and I.J.  Thompson, Ann. Phys. 243, 420 (1995).
\bibitem{Satchler-Love}G.R. Satchler, and W.G. Love, Physics Reports, 55, 183 (1979).
\bibitem{Jim-Jef}J. S. Al-Khalili and J. A. Tostevin, Phys. Rev. Lett. 76, 3903 (1996).
\bibitem{Ron}R. C. Johnson, J. S. Al-Khalili, and J.A. Tostevin, Phys. Rev. Lett. 79, 2771 (1997).
\bibitem{Tan85}  I. Tanihata et. al., Phys. Rev. Lett. 55, 2676 (1985); I. Tanihata et al., Phys. Lett. B160, 380 (1985).
\bibitem{Tan88} I.  Tanihata,  T.  Kobayashi,  O.  Yamakawa,  S.  Shimoura, K.  Ekuni,  K.  Sugimoto,  N.  Takahashi,  T.  Shimoda,  and H. Sato, Phys. Lett. B206, 592 (1988).
\bibitem{AH09} D. M. Andrade and M. S. Hussein, Phys. Rev. C 89, 034610 (2009).
\bibitem{JG00} R. C. Johnson, and C. J. Goebel, Phys. Rev. C 62, 027603 (2000).
\bibitem{BoBe01} A. Bonaccorso, G.F. Bertsch, Phys.Rev. C63, 044604 (2001).
\bibitem{Marg03} J. Margueron, A. Bonaccorso, D.M. Brink, Nucl.Phys. A720, 337 (2003); Erratum Nucl. Phys. A741, 381 (2004).
\bibitem{BBB04}A.Bonaccorso, D.M.Brink, C.A.Bertulani, Phys.Rev. C 69, 024615 (2004)
\bibitem{Ibra05}A.A. Ibraheem, A. Bonaccorso, Nucl.Phys. A748, 414 (2005).
\bibitem{Bla07} G.Blanchon, A. Bonaccorso, D.M. Brink, A.Garcia-Camacho, N.Vinh Mau, Nucl.Phys. A784, 49 (2007).
\bibitem{Kum11} R. Kumar, A. Bonaccorso, Phys.Rev. C 84, 014613 (2011).
\bibitem{Rav12} Ravinder Kumar and Angela Bonaccorso, Phys. Rev. C 86, 061601(2012).
\bibitem{Fla12} F. Flavigny, A. Obertelli, A. Bonaccorso, G. F. Grinyer, C. Louchart, L. Nalpas, and A. Signoracci Phys. Rev. Lett. 108, 252501 (2012).
\bibitem{Bert03} C.A. Bertulani, C.M. Campbell, and T. Glasmacher, Comput. Phys. Commun. 152, 317 (2003).
\bibitem{Bert05} C.A.Bertulani, Phys. Rev. Lett. 94, 072701 (2005).
\bibitem{Baye1}D. Baye, P. Capel, and G. Goldstein, Phys. Rev. Lett. 95, 082502 (2005).
\bibitem{Baye2}G. Goldstein, P. Capel, and D. Baye, Phys. Rev. C 76, 024608 (2007).
\bibitem{Cape10} P. Capel,  M.S. Hussein, and D. Baye, Phys. Lett. B 693, 448 (2010).
\bibitem{OB09} K. Ogata and C.A. Bertulani, Prog. Theor. Phys. (Letter) 121 (2009),
\bibitem{OB10} K. Ogata and C.A. Bertulani, Prog. Theor. Phys. 123, 701 (2010).
\bibitem{Fuk14} T.Fukui, K.Ogata, P.Capel, Phys.Rev. C 90, 034617 (2014).
\bibitem{Min14}K.Minomo, T.Matsumoto, K.Egashira, K.Ogata, M.Yahiro, Phys.Rev. C 90, 027601 (2014).
\bibitem{Vale14} J.A. Lay Valera, A.M.  Moro Mu\~noz, J.M. Arias Carrasco,  Phys. Rev. C  89, 014333 (2014). 
\bibitem{Migu14} J.M. Arias Carrasco, J.A. Lay Valera, A.M.  Moro Mu\~noz, Phys. Rev. C 89, 064609 (2014). 
\bibitem{Garc13}J.P. Fern\'andez Garcia, et al.,  Phys. Rev. Lett. 110, 142701 (2013) 
\bibitem{CDCC1}G. H. Rawitscher, Phys. Rev. C 9, 2210 (1974).
\bibitem{CDCC2}M. Yahiro, Y. Iseri, H. Kameyama, M. Kamimura, and M. Kawai, Prog. Theor. Phys. Suppl. 89, 32 (1986).
\bibitem{CDCC3}N. Austern, Y. Iseri, M. Kamimura, M. Kawai, G. Rawitscher, M. Yahiro, Physics Reports, 154, 125 (1987).
\bibitem{CDCC4} M. Yahiro, T. Matsumoto, K. Minomo, T. Sumi, and S. Watanabe, Prog. Theor. Phys. Supp. 196, 87 (2012).
\bibitem{descouvemont13}P. Descouvemont and M. S. Hussein, Phys. Rev. Lett. 111, 082701 (2013).
\bibitem{descouvemont15}P. Descouvemont, T. Druet, L. F. Canto, and M. S. Hussein, Phys. Rev. C 91, 024606 (2015).
\bibitem{BDH14} C. A. Bertulani, P. Descouvemont, and M. S. Hussein, Eur. Phys. J.  69, 00020 (2014).
\bibitem{BCDFH15} C. A. Bertulani, B. V. Carlson, P. Descouvemont, T. Frederico, and M. S. Hussein, to be published.
\bibitem{Ik88} K. Ikeda, INS Report JHP-7 (1988) (in Japanese); K. Ikeda, Nucl. Phys. A 538 355c (1992).
\bibitem{Auma05} T. Aumann, Eur. Phys. J A 26, 441 (2005). 
\bibitem{Ber07c} C.A. Bertulani, Nucl. Phys. A 788, 366 (2007). 
\bibitem{SAZ13} D. Savran, T. Aumann und A. Zilges, Prog. Part. Nucl. Phys. 70, 210 (2013).
\bibitem{ANa13} T. Aumann, T. Nakamura, Physica Scripta 152 14012 (2013).
\bibitem{SPF13} M. Scheck, V. Y. Ponomarev, M. Fritzsche, J. Joubert, T. Aumann, et al. Phys. Rev. C 88, 44304 (2013).
\bibitem{Abra12}S. Abrahamyan et al., Phys. Rev. Lett. 108, 112502 (2012). 
\bibitem{Hor01}C. J. Horowitz, S. J. Pollock, P. A. Souder, and R. Michaels, Phys. Rev. C 63, 025501 (2001).
\bibitem{Garc92}C. Garcia-Recio, J. Nieves, and E. Oset, Nucl. Phys. A547, 473 (1992).
\bibitem{Ray78}L. Ray, W. R. Coker, and G. W. Hoffmann, Phys. Rev. C 18, 2641 (1978).
\bibitem{Star94}V. E. Starodubsky and N. M. Hintz, Phys. Rev. C 49, 2118 (1994).
\bibitem{Clar03}B. C. Clark, L. J. Kerr, and S. Hama, Phys. Rev. C 67, 054605 (2003).
\bibitem{Trzc01} A. Trzcinska et al., Phys. Rev. Lett. 87, 082501 (2001).
\bibitem{Lens09} H. Lenske, Hyperfine Interact. 194, 277 (2009).
\bibitem{Hus95} M.S. Hussein,  Nucl. Phys. A 588, 85 (1995).
\bibitem{Gom12} P.R.S. Gomes, D.R. Otomar, L.F. Canto, J. Lubian, R. Linares, D.H. Luong, M. Dasgupta, D J. Hinde, M.S. Hussein, J. Phys. G 39, 115103 (2012).
\bibitem{Canto06}L. F. Canto, P. R. S. Gomes, R. Donangelo, and M. S. Hussein, Phys. Reports,  424, 1 (2006).
\bibitem{HT12} K. Hagino, N. Takigawa, Prog. Theor. Phys. 128, 1061  (2012).
\bibitem{Esbensen}B.B. Back, H. Esbensen, C.L. Jiang, and K.E. Rehm, Rev. Mod. Phys. 86, 317 (2014).
\bibitem{Canto15}L. F. Canto, P. R. S. Gomes, R. Donangelo, J. Lubian, and M. S. Hussein, Phys. Reports,  in press (2015).
\bibitem{HCD03} M.S. Hussein, L.F. Canto, R. Donangelo, Nucl. Phy. A 722c, 321 (2003).
\bibitem{IAV} M. Ichimura, N. Austern, and C. M. Vincent, Phys. Rev. C 32, 431 (1985).
\bibitem{HM1} M. S. Hussein and K. W. McVoy, Nucl. Phys. A 445, 124(1985).
\bibitem{HM2}M. S. Hussein, and R. C. Mastroleo, Nucl. Phys. A 491, 468 (1989).
\bibitem{UT} T. Udagawa, and T. Tamura, Phys. Rev. C 33, 494 (1986).
\bibitem{ULT}T. Udagawa, Y. Lee, and T. Tamura, Phys. Lett. B 196, 291 (1987).
\bibitem{HFM} M. S. Hussein, T. Frederico, and R. C. Mastroleo, Nucl. Phys. A 511, 269 (1990).
\bibitem{Ichimura} M. Ichimura, Phys. Rev. C 41, 838 (1990).
\bibitem{Surrogate}Jutta E. Escher, Jason T. Burke, Frank S. Dietrich, Nicholas D. Scielzo, Ian J. Thompson, and Walid Younes, Rev. Mod. Phys. 84, 353 (2012).
\bibitem{Serb47} R. Serber Phys. Rev. 72, 1008 (1947). 
\bibitem{But50} S.T. Butler, Phys. Rev. 80, 1095 (1950); Nature 166, 709 (1950); Proc. Roy. Soc. 208 A, 559 (1951).
\bibitem{Marta} D. Marta, L. F. Canto and R. Donangelo, Phys. Rev. C89, 034625 (2014).
\bibitem{Lei15} J. Lei, and A. M. Moro, to be published in Phys. Rev.   C (2015).
\bibitem{Potel15} G. Potel, F. M. Nunes, and I. J. Thompson, to be   published in Phys. Rev. C (2015). arXiv:1508.04822.
\bibitem{Carlson15} B. V. Carlson, R. Capote, and M. Sin, Few-Body Systems (2015). arXiv:1508.01.01466.
\bibitem{Wigner}E. P.  Wigner, Proc. Cambridge Phil. Soc. 47 70 (1951).
\bibitem{Wigner2}E. P. Wigner, Ann. of Math. 62, 548 (1955).
\bibitem{Dyson}F. J.  Dyson, J. Math. Phys. 3,  379 (1962).
\bibitem{BG84} O. Bohigas, and M. J. Giannoni, ``Chaotic Motion and Random Matrix Theories", Lecture Notes in Physics  209, 1 (1984).
\bibitem{DiAc15}B. Dietz and A. Richter, Chaos. 25, 097601 (2015).
\bibitem{Dietz14} B. Dietz, T. Klaus, M. Miski-Oglu, and A. Richter, Phys. Rev. E 89, 032909 (2014).
\bibitem{Kum13} S. Kumar, A. Nock, H.-J. Sommers, T. Guhr, B. Dietz, M. Miski-Oglu, A. Richter, and F. Sch\"afer, Phys. Rev. Lett. 111, 030403 (2013).
\bibitem{BS1963} D. M. Brink and R. O. Stephen, Phys. Lett. 5 77 (1963).
\bibitem{Ram12} J. Ramos,  A. Barbosa, D. Bazeia, M.S. Hussein, C.  Lewenkopf, Phys. Rev. B 86,  235112 (2012).
\bibitem{HuPa00} M.S. Hussein, M.P. Pato, Physica A 285, 383 (2000).
\bibitem{Weiden1}G. E. Mitchell, A. Richter, and H. A. Weidenm\"uller, Rev. Mod. Phys. 82, 2845 (2010).
\bibitem{Weiden2}B. Dietz, T. Friedrich, H. L. Harney, M. Miski-Oglu, A. Richter, F. Sch\"afer, and H. A. Weidenm\"uller,  Phys. Rev. E 81, 036205 (2010).
\bibitem{HJ87} P.G. Hansen and B. Jonson, Europhys. Lett. 4, 409 (1987).
\bibitem{BeBau88} C.A. Bertulani and G. Baur, Nucl. Phys. A 480, 615 (1988).
\bibitem{fairf} http://www.fair-center.eu
\bibitem{frib} http://www.frib.msu.edu
\bibitem{Zhu93} M.V. Zhukov, et. al., Phys. Reports 231, 151 (1993).
\bibitem{Yam06} M.T. Yamashita, T. Frederico,  M.S. Hussein, Mod. Phys. Lett. A 21, 1749 (2006).
\bibitem{Efi70} V. Efimov, Phys. Lett. B33, 563 (1970).
\bibitem{Li08} Bao-An  Li, L.W. Chen, and C.M. Ko, Phys. Rep. 464, 113 (2008).
\bibitem{BaLi00} Bao-An Li, Phys. Rev. Lett. 85, 4221 (2000).
\bibitem{BaLi02} Bao-An Li, Phys. Rev. Lett. 88, 192701 (2002).
\bibitem{Chen05} Lie Wen Chen, Che Min Ko, Bao-An Li, Phys. Rev. Lett. 94, 032701 (2005).
\bibitem{Kra06} T. Kraemer, et al., Nature 440, 7082 (2006).
\bibitem{BH07} C. A. Bertulani, M. S. Hussein, Phys. Rev. C76, 051602 (2007).
\bibitem{Ste13} D. Steppenbeck, et. al., Nature 502, 207 (2013).
\bibitem{Ots05} T. Otsuka, et. al., Phys. Rev. Lett. 95, 232502 (2005).
\bibitem{Suz03} T. Suzuki, R. Fujimoto  and T. Otsuka,  Phys. Rev. C 67 044302 (2003).
\bibitem{Hon02}M. Honma T. Otsuka, B.A. Brown and T. Mizusaki, Phys. Rev. C 65 061301 (2002); ibid. 69, 034335 (2004).
\bibitem{Otsu10} T. Otsuka, T. Suzuki, M. Honma, Y. Utsuno, N. Tsunoda, K. Tsukiyama, and M. Hjorth-Jensen, Phys. Rev. Lett. 104, 012501 (2010).
\bibitem{Yua12} C. Yuan, T. Suzuki, T. Otsuka, F. Xu and N. Tsunoda, Phys. Rev. C 85, 064324 (2012).
\bibitem{Suz13} T. Suzuki and M. Honma, Phys. Rev. C 87, 014607 (2013).
\bibitem{Cot10} P.D. Cottle, Nature 465, 430 (2010).
\bibitem{Sch94} R. Schneider, et. al., Z. Phys. A 348, 241 (1994).
\bibitem{Iked68} K. Ikeda, N. Takigawa and H. Horiuchi, Prog. Theor. Phys. Suppl. (extra number), 464 (1968).
\bibitem{Feld98} H. Feldmeier, T. Neff, R. Roth, J. Schnack, Nucl. Phys. A632, 61 (1998).
\bibitem{Itaga00} N. Itagaki, S. Okabe, Phys. Rev. C 61, 044306 (2000).
\bibitem{Free01} M. Freer, et al.,  Phys. Rev. C 63 034301 (2001).
\bibitem{DeBa04} P. Descouvemont  and D. Baye,  Phys. Lett. B 505, 71 (2001).
\bibitem{Beta02} R. Id Betan, R.J. Liotta, N. Sandulescu and T. Vertse, Phys. Rev. Lett. 89, 042501 (2002).
\bibitem{Neff03} T. Neff, H. Feldmeier, Nucl. Phys. A713, 311 (2003).
\bibitem{Okolo03} J. Okolowicz, M. Ploszajczak and I. Rotter, Phys. Rep. 374, 271 (2003).
\bibitem{Voly06} A. Volya and V. Zelevinsky, Phys. Rev. C74, 064314 (2006).
\bibitem{Mic07}N. Michel, W. Nazarewicz and M. Ploszajczak, Phys. Rev. C75, 031301(2007); Nucl. Phys. A794, 29 (2007).
\bibitem{Mas07} H. Masui, N. Itagaki, Phys. Rev. C 75, 054309 (2007).
\bibitem{Itaga07} N. Itagaki, M. Kimura, C. Kurokawa, M Ito, W von Oertzen, Phys. Rev. C 75, 037303 (2007).
\bibitem{Ito08} M. Ito, N. Itagaki, H. Sakurai and K. Ikeda, Phys. Rev. Lett. 100 182502 (2008).
\bibitem{Roth10}R. Roth, T. Neff, H. Feldmeier, Prog. Part. Nucl. Phys. 65, 50 (2010).
\bibitem{Oko13} J. Okolowicz, M. Ploszajczak, W. Nazarewicz, Prog. Theor. Phys. Supp. 196, 230 (2012).
\bibitem{Free12} M. Freer, et al., Phys. Rev. C 86, 034320 (2012).
\bibitem{Beri12}J. Beringer, et al., ``Review of Particle Physics", Phys. Rev. D 86 010001 (2012).
\bibitem{Gar14}Alberto Garfagnini, Int. J. Mod. Phys. Conf. Ser., 31, 1460286 (2014).
\bibitem{Cau08} E. Caurier, J. Menendez, F. Nowacki and A. Poves,  Phys. Rev. Lett. 100, 052503 (2008).
\bibitem{Men09} J. Menendez, A. Poves, E. Caurier, and F. Nowacki, Nucl. Phys. A 818, 139 (2009.
\bibitem{Rod06} V. A. Rodin, A. Faessler, F. Simkovic and P. Vogel, Nucl. Phys. A 766, 107 (2006), Erratum-ibid. 793, 213, (2007).
\bibitem{Kort07}M. Kortelainen, O. Civitarese, J. Suhonen and J. Toivanen,  Phys. Lett. B 647, 128 (2007).
\bibitem{Bare13} J. Barea, J. Kotila and F. Iachello, Phys. Rev. C 87, 014315 (2013). 
\bibitem{Pine10}T. R. Rodriguez and G. Martinez-Pinedo, Phys. Rev. Lett. 105, 252503 (2010).
\bibitem{Vaq13}N. L\'opez Vaquero, T. R. Rodriguez and J. L. Egido,  Phys. Rev. Lett. 111, 142501 (2013).
\bibitem{BB88} C.A. Bertulani and G. Baur, Phys. Reports 163, 299 (1988).
\bibitem{BG10}  C.A. Bertulani and A. Gade, Phys. Rep. 485, 195 (2010).
\bibitem{Tribb14}R.Tribble, C.A. Bertulani, M. La Cognata, A. Mukhamedzhanov, and C. Spitaleri, Rep. Prog. Phys. 77, 106901 (2014). 
\bibitem{BP02} ``Neutron Star Crust", C.A. Bertulani and J. Piekarewicz, editors,, Nova Science, Hauppage, NY (2012),
\bibitem{Mas15} Masayuki Matsuo, Phys. Rev. C 91, 034604 (2015).
\bibitem{Brow00} Alex Brown, Phys. Rev. Lett. 85, 5296 (2000).
\bibitem{Dan02} P Danielewicz, R Lacey, WG Lynch, Science 298, 1592 (2002).
\bibitem{Ber07a} C.A. Bertulani, J. Phys. G 34, 315 (2007).
\bibitem{Cen10} M. Centelles, X. Roca-Maza, X. Vi\~nas, and M. Warda, Phys. Rev. C 82, 054314 (2010).
\bibitem{Wie09} O. Wieland et al., Phys. Rev. Let.. 102, 092502 (2009).
\bibitem{Ros13} D. Rossi et al., Phys. Rev. Lett. 111, 242503 (2013). 
\bibitem{SAZ11} D. Savran, T. Aumman and A. Zilges, Prog. Part. Nucl. Phys. 70, 210 (2013).  
\bibitem{Ber07b} C.A. Bertulani, Phys. Rev. C 75, 024606 (2007).
\bibitem{Bru68} Keith A. Brueckner, Sidney A. Coon, and Janusz Dabrowski, Phys. Rev. 168, 1184 (1968).
\bibitem{TB01} S. Typel and B.A. Brown, Phys. Rev. C 64, 027302 (2001).
\bibitem{Furn02} R.J. Furnstahl, Nuc. Phys. A 706, 85 (2002).
\bibitem{Go98} S. Goriely, Phys. Lett. B 436, 10 (1998).
\bibitem{BHR03} M. Bender, P.-H. Heenen, and P.-G. Reinhard, Rev. Mod. Phys. 75, 121 (2003).
\bibitem{RN10} P.-G. Reinhard and W. Nazarewicz, Phys. Rev. C 81, 051303 (2010).
\bibitem{Piek11} J. Piekarewicz, Phys. Rev. C 83, 034319 (2011).
\bibitem{Tam11}A. Tamii et al., Phys. Rev. Lett. 107, 062502 (2011).
\bibitem{Pol12} I. Poltoratska et al., Phys. Rev. C 85, 041304 (2012). 
\bibitem{Krum15}A.M. Krumbholz, et al., Phys. Lett. B 744, 7 (2015).
\bibitem{CFH13} C.A. Bertulani, J. Fuqua and M.S. Hussein, Astrophys. J. 767, 67 (2013).
\bibitem {Ber88} C.A.Bertulani  and G.Baur, Phys. Reports 163, 299 (1988).
\bibitem {BBR86}G. Baur, C. Bertulani and H. Rebel, Nucl. Phys. A459, 188 (1986).
\bibitem{Tohru} T. Motobayashi, et al., Phys. Rev. Lett. 73, 2680 (1994).
\bibitem{Izs13} R. Izsak, et al., Phys. Rev. C 88, 065808 (2013). 
\bibitem {EBS05}H. Esbensen, G.F. Bertsch, and K. Snover, Phys. Rev. Lett. 94,042502 (2005); Moshe Gai, Phys. Rev. Lett. 96, 159201 (2006); H. Esbensen, G.F. Bertsch, and K. Snover, Phys. Rev. Lett. 94, 042502 (2005).
\bibitem{BG98} C.A. Bertulani and M. Gai, Nucl. Phys. A636,  227 (1998).
\bibitem{BC92} C.A. Bertulani and L.F. Canto, Nucl. Phys. A 539, 163 (1992).
\bibitem {BBK92} G. Baur, C.A. Bertulani and D.M. Kalassa, Nucl. Phys. A550, 527 (1992).
\bibitem {BB93} G.F. Bertsch and C.A. Bertulani, Nucl. Phys. A556, 136 (1993); Phys. Rev. C49, 2834 (1994); H. Esbensen, G.F. Bertsch and C.A. Bertulani, Nucl. Phys. A581, 107 (1995).
\bibitem {GB95} M. Gai and C.A. Bertulani, Phys. Rev. C52, 1706 (1995); Nucl. Phys. A636, 227 (1998).
\bibitem {EB96} H. Esbensen and G. Bertsch, Nucl. Phys. A600, 37 (1996).
\bibitem {Ber94} C.A. Bertulani, Phys. Rev. C49, 2688 (1994); Nucl. Phys. A587, 318 (1995); Z. Phys. A356, 293 (1996).
\bibitem {Ber02} C.A. Bertulani, Phys. Lett. B3, 205 (2002).
\bibitem{Rob73} R.G.H. Robertson, Phys. Rev. C 7, 543 (1973).
\bibitem{WK81}R. D. Williams and S. E. Koonin, Phys. Rev. C 23, 2773 (1981).
\bibitem{EB95} H. Esbensen and G. Bertsch, Phys. Lett. B 359, 13 (1995).
\bibitem{BB00} D. Baye and E. Brainis, Phys. Rev. C 61, 025801 (2000).
\bibitem{DT03} B. Davids and S. Typel, Phys. Rev. C 68, 045802 (2003).
\bibitem{Jun10} Junting Huang, C.A. Bertulani and V. Guimaraes, Atomic Data and Nuclear Data Tables 96, 824 (2010). 
\bibitem{RH11} G. Rupak and R. Higa, Phys. Rev. Lett. 106, 222501 (2011).
\bibitem{ZNP14a} Xilin Zhang, Kenneth M. Nollett, and D. R. Phillips, Phys. Rev. C 89, 024613 (2014).
\bibitem{ZNP14b} Xilin Zhang, Kenneth M. Nollett, and D. R. Phillips, Phys. Rev. C 86, 051602(R) (2014).
\bibitem{TLT78} Y.C. Tang, M. LeMere and D.R. Thompsom, Phys. Rep. 47, 167 (1978).
\bibitem{DB94} P. Descouvemont, D. Baye, Nucl. Phys. A 567, 341 (1994).
\bibitem{HW53} D.L. Hill and J.A. Wheeler, Phys. Rev. 89,1102 (1953).
\bibitem{LK85} K. Langanke, S. Koonin: Nucl. Phys. A439, 384 (1985); A439. 384 (1985).
\bibitem{WW93} T.A. Weaver and S.E. Woosley, Phys. Rep. 227, 65 (1993).
\bibitem{Nol01} Kenneth M. Nollett, Phys. Rev. C 63, 054002 (2001).
\bibitem{NBC06} P. Navr\'atil, C.A. Bertulani, and E. Caurier, Phys. Lett. B 634 (2006) 191; Nucl. Phys. A787, 539 (2007).
\bibitem{QN08} Sofia Quaglioni and Petr Navr\'atil, Phys. Rev. Lett.  101, 092501 (2008).
\bibitem{NRQ11} P. Navratil, R. Roth and S. Quaglioni, Phys. Lett. B 704, 379 (2011).
\bibitem{NQ12} Petr Navratil, Sofia Quaglioni, Phys. Rev. Lett.108, 042503 (2012).
\bibitem{Bau86} G. Baur, Phys. Lett. B 178, 135 (1986).
\bibitem{Spi11} C. Spitaleri, A. M. Mukhamedzhanov, L. D. Blokhintsev, M. La Cognata, R. G. Pizzone, A. Tumino, Physics of Atomic Nuclei 74,  1725  (2011). 
\bibitem{Mukunp} A. M. Mukhamedzhanov (unpublished).
\bibitem{Che96} S. Cherubini et al., Astrophys. J. 457, 855 (1996).
\bibitem{Zad89} M. Zadro, D. Miljanic, C. Spitaleri, G. Calvi, M. Lattuada, and F. Riggi, Phys. Rev. C 40, 181 (1989).
\bibitem{Spi01} C. Spitaleri et al., Phys. Rev. C 63, 055801 (2001).
\bibitem{Lat01} M. Lattuada et al., Astrophys. J. 562, 1076 (2001).
\bibitem{Piz13} G. Pizzone,et al., Phys. Rev. C 87, 025805 (2013). 
\bibitem{Spi13} Claudio Spitaleri, J.  Phys.: Conference Series 420,  012137 (2013).
\bibitem{Piz14} R.G. Pizzone, et. al., Astrophys. J. 786, 112  (2014).
\bibitem{Xu94} H.M. Xu, C.A. Gagliardi, R.E. Tribble, A.M. Mukhamedzhanov and N.K. Timofeyuk, Phys. Rev. Let. 73, 2027 (1994).
\bibitem{Mu10} A.M. Mukhamedzhanov et al, J. Phys. Conference Series 202,  012017  (2010).
\bibitem{Con07} M. La Cognata et al., Phys. Rev. C 76, 065804 (2007).
\bibitem{Tri06} R.E. Tribble et al., Proceedings of Science (online at: http://pos.sissa.it/index.html) (2007).
\bibitem{Muk90}  A.M. Mukhamedzhanov and N.K. Timofeyuk, JETP Lett. 51, 282 (1990).
\bibitem{CB70} J.D. Cramer and H.C. Britt, Phys. Rev. C 2, 2350 (1970).
\bibitem{ED10} J.E. Escher and F.S. Dietrich, Phys. Rev. C 81, 024612 (2010).
\bibitem{CI10} S. Chiba and O. Iwamoto, Phys. Rev. C 81, 044604 (2010).
\bibitem{Kes10} G. Kessedjian et al., Phys. Lett B 692, 297 (2010).
\bibitem{AGS67} P. Grassberger and W. Sandhas, Nucl. Phys. B 2,  181 (1967); E.O. Alt, P. Grassberger, W. Sandhas, Phys. Rev. C 1, 85 (1970).
\bibitem{Muk14} A. M. Mukhamedzhanov, D. Y. Pang, C. A. Bertulani, and A. S. Kadyrov, Phys. Rev. C 90, 034604 (2014).
\bibitem{DF09} A. Deltuva and A. C. Fonseca, Phys. Rev. C 79, 014606 (2009).
\bibitem{Sas12} M. Sasano, et. al.,  Phys. Rev. C 86, 015809 (2012).
\bibitem{MLD00} G. Martinez-Pinedo, K. Langanke and D. Dean, Astrop. J. Sup. 126, 493 (2000). 
\bibitem{Taddeucci1987} T.N. Taddeucci, et al., Nucl.~Phys.~A 469, 125 (1987).
\bibitem {Ber93} C.A. Bertulani, Nucl. Phys. A 554, 493 (1993); C.A. Bertulani and P. Lotti, Phys. Lett. B 402, 237 (1997); C.A. Bertulani and D. Dolci, Nucl. Phys. A674, 527 (2000).
\bibitem{Zegers2007} R.G.T. Zegers, et al., Phys. Rev. Lett. 99, 202501 (2007). 
\bibitem{HW04} R.C. Hwa and X.N. Wang, eds., ``Quark-Gluon Plasma 3", Singapore: World Scientific (2004).
\bibitem{Kog83} B. Kogut, M. Stone, H. W. Wyld, W. R. Gibbs, J. Shigemitsu, S. H. Shenker and D. K. Sinclair, Phys. Rev. Lett. 50, 393 (1983).
\bibitem{Baz13} A. Bazavov, J. Phys. Conf. Ser. 446, 012011 (2013).
\bibitem{Fodo09}Z. Fodor and S. D. Katz, ``The phase diagram of quantum chromodynamics", Landolt-Boernstein, Vol. 1-23A (2009). arXiv:0908.3341.
\bibitem{Miya95}O. Miyamura, Phys. Lett. B 353, 91 (1995).
\bibitem{Wol95}R. M. Woloshyn, Phys. Rev. D 51, 6411 (1995).
\bibitem{Agg00} M.M. Aggarwal et al., Phys. Rev. Lett. 85, 3595 (2000).
\bibitem{Bat06} S. Bathe et al., Nucl. Phys. 774, 103 (2006).
\bibitem{Dam06} S. Damjanovic et al., Nucl. Phys. A 774, 715 (2006).
\bibitem{Ada06} A. Adare et al., nucl-ex/06011020 (2006).
\bibitem{Bue06} H. Buesching et al., Nucl. Phys. 774, 103 (2006).
\bibitem{GB02}V. P. Goncalves and C. A. Bertulani, Phys. Rev. C 65, 054905 (2002).
\bibitem{BKN05} C. A. Bertulani, S. R. Klein, and J. Nystrand,  Ann. Rev. Nucl. Part. Sci. 55, 271 (2005).
\bibitem{Bal08} A. Baltz,  et al., Phys. Rep. 458 (2008).
\bibitem{AB11} Adeola Adeluyi and Carlos A. Bertulani, Phys. Rev. C 84, 024916 (2011); Phys. Rev. C 85, 044904 (2012); Phys. Rev. C 86, 047901 (2012). 
\bibitem{AN13} A. Adeluyi and T. Nguyen,  Phys.Rev. C87, 027901 (2013).
\bibitem{Reby12} V. Rebyakova, M. Strikman, and M. Zhalov,  Phys. Lett. B710,  647 (2012).
\bibitem{Abe13} B. Abelev et al.,  Phys. Lett. B 718, 1273 (2013).
\bibitem{expups} ALICE collaboration, arXiv:1508.05076 [nucl-ex] , CERN-PH-EP-2015-156 (2015).
\bibitem{SZ86} E.V. Shuryak and O.V. Zhirov, Phys. Lett. B 171, 99 (1986).
\bibitem{CERN00} http://web.cern.ch, news (2000).
\bibitem{Gyu04} Miklos Gyulassy, ``The QGP Discovered at RHIC". arXiv:nucl-th/0403032 (2004).
\bibitem{BNL10} https://www.bnl.gov, news (2010)
\bibitem{Nat12} http://blogs.nature.com, news (2012).
\bibitem{IV03} E. Iancu, R. Venugopalan, hep-ph/0303204 (2003).
\bibitem{LM06}T. Lappi and L. McLerran, Nucl. Phys. A 772, 200 (2006).
\bibitem{GV06} F. Gelis, R. Venugopalan, Acta Phys. Polon. B 37, 3253 (2006).
\bibitem{CMS12} CMS collaboration,  Phys. Lett. B 716, 30 (2012). arXiv:1207.7235. 
\bibitem{ATLAS12} ATLAS collaboration,  Phys. Lett. B 716, 1 (2012). arXiv:1207.7214. 
\bibitem{GS71} J.-L. Gervais and B. Sakita,  Nucl. Phys. B 34 , 632 (1971). 
\bibitem{Vol72}  D.V. Volkov, V.P. Akulov, Pisma Zh. Eksp. Teor. Fiz. 16, 621 (1972);  Phys. Lett. B 46 (1973) 109; V.P. Akulov and D.V. Volkov, Teor. Mat. Fiz. 18, 39 (1974).
\bibitem{Ram71} P. Ramond,  Phys. Rev. D 3, 2415 (1971). 
\bibitem{WZ74} J. Wess and B. Zumino, Nucl. Phys. B 70, 39 (1974).
\bibitem{Mar97} Stephen P. Martin,  ``A Supersymmetry Primer". arXiv:hep-ph/9709356 (1997).
\bibitem{Kal21} Theodor Kaluza,  Sitzungsber. Preuss. Akad. Wiss. Berlin. (Math. Phys.): 966, 1 (1921).
\bibitem{Ark98} N. Arkani-Hamed, S. Dimopoulos, G. Dvali, Phys. Lett. B 429, 263. arXiv:hep-ph/9803315;  Phys. Rev. D 59,  086004 (1998).
\bibitem{App01}Thomas Appelquist, Hsian-Chia Cheng, Bogdan A. Dobrescu,  Phys. Rev. D 64, 035002 (2001).
\bibitem{Rizz01} Thomas G. Rizzo, Phys. Rev. D 64, 095010 (2001). arXiv:hep-ph/0106336. 
\bibitem{McM02} C. Macesanu, C.D.  McMullen, S.  Nandi,  Phys. Rev. D 66, 015009 (2002). arXiv:hep-ph/0201300.  
\bibitem{isolde} http://isolde.web.cern.ch 
\bibitem{Bau96} G. Baur, et al.,  Phys. Lett. B 368,  251 (1996).
\bibitem{BeBa88} C.A.Bertulani and G. Baur,  Braz. J. Phys. 18, 559 (1988).
\bibitem{Blan97} G. Blanford, et al., Phys. Rev. Lett.  80,  3037 (1997).
\bibitem{BeBa98} C.A. Bertulani and G. Baur,  Phys. Rev. D58, 034005 (1998).
\bibitem{AL15} ALICE collaboration, Nature Physics Lett. doi:10.1038/nphys3432 (2015).
\bibitem{ITER} https://www.iter.org
\bibitem{Suda} Toshimi Suda et al., Prog. Theor. Exp. Phys.  03C008 (2012).
\bibitem{Raju15} Raju Venugopalan, ``Why we need an Electron-Ion Collider", arXiv:1507.08128 [nucl-th] (2015).
\bibitem{Abel12} J.L. Abelleira Fernandez, et al., ``LHeC Study Group Collaboration", J. Phys. G39, 075001 (2012).

\end{thebibliography}
\end{document}